%% file: newZp.tex
\documentclass[a4paper,12pt]{article}
\pdfoutput=1
\usepackage{graphicx}
\usepackage{amsfonts}
\usepackage{amsmath}
\usepackage{multirow}
\usepackage{mathrsfs}
\usepackage{amssymb}
\usepackage{verbatim}
\usepackage{float}
\usepackage{url,tabularx,amsfonts,slashed,amsmath,array}
\usepackage{sidecap}

\newcommand{\be}{\begin{eqnarray}}
\newcommand{\ee}{\end{eqnarray}}

\newcommand{\notB}{\slash\!\!\!\!\!\:B}
\newcommand{\notL}{\slash\!\!\!\!\!\!\;\:L}

\begin{document}

\begin{flushright}
preprint SHEP-12-15\\
preprint FR-PHENO-2012-016\\
\today
\end{flushright}
\vspace*{1.0truecm}

\begin{center}
{\large\bf Non-exotic $Z'$ signals in $\ell^+\ell^-$, $b\bar b$ and $t\bar t$ final states at the LHC}\\
\vspace*{1.0truecm}
{\large L. Basso$^1$, K. Mimasu$^2$ and S. Moretti$^{2,3}$}\\
\vspace*{0.5truecm}
{\it $^1$Physikalisches Institut, Albert-Ludwigs-Universit\"at Freiburg\\
D-79104 Freiburg, Germany}\\
\vspace*{0.25truecm}
{\it $^2$School of Physics \& Astronomy, University of Southampton, \\
Highfield, Southampton, SO17 1BJ, UK}\\
\vspace*{0.25truecm}
{\it $^3$Particle Physics Department\\ Rutherford Appleton Laboratory\\
Chilton, Didcot, Oxon OX11 0QX, UK}\\
\end{center}

\vspace*{1.0truecm}
\begin{center}
\begin{abstract}
\noindent
In the attempt to fully profile a $Z'$ boson accessible at the Large Hadron Collider (LHC),
we study the sensitivity of di-lepton (for the electron, muon and tauon cases) and di-quark (for the case of the heavy flavours, $t$ and, possibly, $b$) samples to the nature of the new gauge state, for a one-dimensional class of non-exotic $Z'$ bosons. Assuming realistic final state reconstruction efficiencies and error estimates, we find that, depending on the CERN collider energy and luminosity, the best chances of extracting the $Z'$ quantum numbers
occur when two or more of these channels are simultaneously explored, as none of them separately enables one to fully probe the parameter spaces of the aforementioned models. Effects of Standard Model (SM) background as well interferences between this and the various $Z'$ signals have been accounted for. A complete study of cross sections and asymmetries (both spatial and spin ones) makes clear the need for complementarity, especially for {their disentanglement over the full parameter space.}
\end{abstract}
\end{center}

\section{Introduction}
\label{sect:intro}

So-called $Z'$ bosons, heavier versions of the SM $Z$ gauge boson,
appear in many Beyond the SM (BSM) scenarios: e.g., in Little Higgs models~\cite{Schmaltz:2005ky},
in frameworks with extra spacial dimensions~\cite{Hewett:2002hv} as well as in
Grand Unified Theories (GUTs) such as $SO(10)$~\cite{mohapatra} (e.g., in Left-Right symmetric models) and
$E_6$~\cite{Hewett:1988xc} types. Furthermore, in
some realisations of Supersymmetry (SUSY) \cite{Chung:2003fi} and in Hidden Valley
models~\cite{Strassler:2006im}, they manifest themselves rather subtly, as messengers between the
SM and the hidden sectors corresponding to such scenarios.

Such $Z'$ states are best searched for at hadron colliders through the Drell-Yan process, i.e., a di-lepton signature proceeding via
$pp(\bar p) \to (\gamma,Z,Z') \to \ell^+\ell^-$, where $\ell=e,\mu$. In fact, the most stringent limits on $Z'$s at both the Tevatron and LHC emerge from these final states. Currently, assuming a sequential
$Z'$, it should be noted that the Tevatron places limits on $M_{Z'}$, at around 1 TeV
\cite{Tevatron} while the LHC sets about 2.3 TeV \cite{LHC}. Recall that a sequential $Z'$ is a state with
generic mass and same coupling to the SM particles as the $Z$ boson, so that limits in the aforementioned models must be obtained by rescaling the results for such a $Z'$ boson\footnote{However, notice that this implicitly assumes that the $Z'$ cannot decay into additional extra matter present in the model spectrum. We will dwell on this aspect later on in the paper.}.
While the above DY signature is clean experimentally and theoretical uncertainties are well under control, it is by no means the only one accessible at the CERN machine. In particular, heavy quark final states, i.e., those accessible via $pp(\bar p) \to (\gamma,Z,Z') \to Q\bar Q$, where $Q=b,t$, also yield detectable rates. Further, they can both be used to profile the $Z'$ boson (that is,
to measure its quantum numbers), alongside the DY channel, at the LHC (see, e.g., \cite{Papaefstathiou:2011kd}). The case has been made recently for $Q=t$ by ourselves~\cite{Basso:2012sz}, building on previous results \cite{Zp-tt}. Here, one exploits the fact that the top (anti)quark pairs decay before hadronising, so that their spin properties are effectively transmitted to their decay products.
This enables one in turn to access the vector and axial couplings of the $Z'$ from which the pair originates.  Further, recall that the electromagnetic charge of the top can be tagged (via a lepton and/or a $b$-jet) \cite{tt-pol}, thanks to which one can define
suitable spatial/spin asymmetries which turn out to be particularly effective
to pin down said couplings of the new gauge boson.
On the same footing, $\tau$ leptons are also short lived so that their spin information is imprinted in their decay product kinematics~\cite{TAUPOL}. Similarly to top quarks, $\tau$ polarisation can be measured by means of its one-prong decays and analyses exist at the LHC already~\cite{TAULHC}.

The case of $Q=b$ brings further complications. 
The displaced vertex tagging allows for the charge measurement of the $b$-quark in the jet, which is needed in the definition of the spatial asymmetries in $b$-quark final states, which therefore seem feasible quantities to deal with. Here the main issue comes from the $b$-tagging performances at high $p_T$ which we discuss in section~\ref{subsubsec:effs}. Nonetheless, one might hope the tagging efficiency to improve due to upgrades in the micro-vertex detectors at both ATLAS and CMS for the $\sqrt{s}=14$ TeV run (see, e.g., \cite{PoS(VERTEX)}). Regarding spin asymmetries, they certainly can be defined also for $b$-quarks. If the $b$-quark hadronises before decaying, it has also been shown that its spin informations can be transmitted to $b$-hadrons (e.g., $\Lambda_b$), albeit with some dilution factors~\cite{BPOL}.  Then, the semi-leptonic decays of the latter would enable spin measurements, as pioneered at LEP~\cite{LEP-b}. It is not clear however if the same measurement is possible at the LHC, due to the higher boost of the emerging lepton and the narrower $b$-jet cone. Despite being nowadays technically challenging and still uncertain for the very near future, as better vertex detectors will be available, it is still interesting to study spin asymmetries for $b$-quarks in view of when such measurements will be available. We will therefore present them in the spirit of suggested observables.


It is the purpose of this paper to compare the yield of $\ell^+\ell^-$ ($\ell = e/\mu, \tau$), $b\bar b$ and $t\bar t$ final states produced at the LHC in presence of a $Z'$ boson and to assess the machine ability to profile the latter in both standard kinematic variables as well as spatial/spin asymmetries, by continuously scanning the parameter space of a one-dimensional class of $Z'$ models, defined in terms of the $Z'$ coupling and mixing parameters, over which we will define benchmark points and lines amenable to experimental investigation.
In doing so we will borrow some (but not all) of the standard $Z'$ benchmarks from, e.g.,  \cite{Accomando:2010fz} and \cite{Basso:2011na},
as well as define new ones.

The plan of the paper is as follows. In the next section we describe the model, outline
our calculation and define the observables to be studied. In sect.~\ref{sec:results} we report and comment on our results. Sect.~\ref{sec:summary} finally presents our conclusions and outlook.

\section{Framework \label{sec:calculation}}
We describe here the model we studied, fixing conventions and relevant features, the details of the code used and the variables that have been analysed.
\input{sect2.tex}

\section{Results \label{sec:results}}
We start by presenting the most up-to-date exclusions from direct searches, from the CMS analysis at $\sim 5$ fb$^{-1}$
\cite{LHC}, followed by a description of the features pertaining to each possible final state, i.e., BRs, cross sections at the LHC for $\sqrt{s}=14$ TeV (as the scope of 7 and 8 TeV stages of the CERN machine is rather minimal for our purposes) and total $Z'$ width. Notice that ATLAS has also published an analysis for $\sim5$ fb$^{-1}$~\cite{LHC}, but their limits are less tight than the CMS ones. Therefore, we will not present them here.
\input{sect3.tex}

\section{Conclusions \label{sec:summary}}
\input{sect4.tex}

\section*{Acknowledgements}
We are grateful to A. Belyaev for the suggestion of the $p_T(b)$ cut and to I.R.~Tomalin for pointing to us the LEP measurements of the polarisation of b-hadrons. LB thanks Stan Lai for helpful discussions about taus. Finally we would like to thank the anonymous referee for referring us to~\cite{CMS-11-008} and additionally stimulating a great deal of internal discussion and development of our understanding of polarisation asymmetries and their role in investigating the $Z^{\prime}$ couplings.
The work of KM and SM is partially supported through the NExT Institute. LB is supported by the
Deutsche Forschungsgemeinschaft through the Research Training Group grant
GRK\,1102 \textit{Physics of Hadron Accelerators}.

\newpage

\end{document}

%% file: sect2.tex
\subsection{The minimal $Z'$ model}
The general class of models we are going to study is the one defined by the so-called non-exotic minimal $Z'$ models~\cite{Basso:2011na,delAguila:1995rb,Carena:2004xs,Chankowski:2006jk,Ferroglia:2006mj,Salvioni:2009mt}. Following the convention of Ref.~\cite{Basso:2011na}, we summarise here just the relevant parts, i.e., the gauge and neutrino sectors, and we refer the reader to the latter publication for a complete description of the model.

When the global $U(1)_{B-L}$ symmetry of the SM is promoted to local, the new $Z'$ gauge boson will mix with the SM $Z$ boson.
In the field basis in which the gauge kinetic terms are diagonal, the covariant derivative is\footnote{In all generality, Abelian field strengths tend to mix. The kinetic term can be diagonalised with a $GL(2,R)$ transformation, leading to the form of the covariant derivative in eq.~(\ref{cov_der}), where the mixing between the two $U(1)$ factors becomes once again evident \cite{delAguila:1995rb,Chankowski:2006jk}.}:
\begin{equation}\label{cov_der}
D_{\mu}\equiv \partial _{\mu} + ig_S T^{\alpha}G_{\mu}^{\phantom{o}\alpha} 
+ igT^aW_{\mu}^{\phantom{o}a} +ig_1YB_{\mu} +i(\widetilde{g}Y + g_1'Y_{B-L})B'_{\mu}\, .
\end{equation}

In our bottom-up approach, we will not require gauge unification at some specific, yet arbitrary, energy scale, fixing the conditions for the extra gauge couplings at that scale. Therefore, in this model, the gauge couplings $\widetilde{g}$ and $g'_1$ are free parameters. To better understand their meaning, let us focus on eq.~(\ref{cov_der}).
The last term of the covariant derivative can be re-written defining an effective coupling $Y^E$ and an effective charge $g_E$:
\begin{equation}\label{eff_par}
g_E Y^E \equiv \, \widetilde{g}Y + g_1'Y_{B-L}.
\end{equation}
As any other parameter in the Lagrangian, $\widetilde{g}$ and $g_1'$ are running parameters~\cite{delAguila:1995rb,Chankowski:2006jk,Ferroglia:2006mj,delAguila:1988jz}, therefore their values ought to be defined at some scale. A discrete set of popular $Z'$ models (see, e.g., Refs.~\cite{Carena:2004xs,Appelquist:2002mw}) can be recovered by a suitable definition of both $\widetilde{g}$ and $g_1'$.

Furthermore, three right-handed neutrinos are required by the anomaly-free conditions, naturally implementing a type-I seesaw mechanism via Yukawa interactions with the SM Higgs field and the new Higgs field needed for the $U(1)_{B-L}$ symmetry breaking to give the $Z'$ boson a mass. A general feature of the seesaw mechanism is that the mass eigenstates, called ``light'' ($\nu_l$) and ``heavy'' ($\nu_h$) neutrinos, are Majorana particles. Because of this, neutrinos only have axial couplings to the neutral gauge bosons. One of the aim of this paper is to investigate the impact of such heavy neutrinos on the observables under study. We will therefore restrict ourselves to two opposite scenarios: a ``decoupled'' case, with heavy neutrinos much heavier than (half of the) $Z'$ mass, thereby disallowing (even off-shell) $Z'\to \nu_h \nu_h$ decays, and a ``very light'' case, where heavy neutrinos are much lighter than the $Z'$ itself (e.g., $m_{\nu_h}=50$ GeV, compatible
with LEP limits). All possible intermediate cases will therefore lie somewhere in between.

\subsection{Structure of the chiral couplings}
To study the asymmetries it is important to understand the chiral structure of the couplings of the $Z'$ gauge boson to fermions, determined by the covariant derivative of eq.~(\ref{cov_der}).

Due to the mixing between the $Z$ and $Z'$ gauge bosons, the couplings of the $Z'$ to fermions are a function of $g'_1$ and $\widetilde{g}$ (notice that also $\theta '$, the angle parametrising the  $Z$-$Z'$ mass mixing, is function of the gauge couplings). In all generality, such couplings can be separated into vector and axial components, $C^f_V(g'_1,\widetilde{g})$ and $C^f_A(g'_1,\widetilde{g})$, respectively, with an interaction term:
\begin{equation}
\mathscr{L}_{Z'}=Z'_\mu\overline{f}\gamma ^\mu \left( C^f_V(g'_1,\widetilde{g}) + \gamma ^5 \, C^f_A(g'_1,\widetilde{g}) \right) f\, ,
\end{equation}
in which we have re-absorbed the gauge couplings in the definition of the vector and axial components.

Because of universal couplings, we can render explicit the chiral structure of the $Z'$ couplings to up-type quarks, down-type quarks and to charged leptons, independently of the fermion generation, as well as the partial decay widths in the approximation of a massless final state.

{\bf Up-type quarks ($u$):}
\begin{eqnarray}\label{up-CV}
C^u_V(g'_1,\widetilde{g}) &=& \frac{-C'\, C_W\, (4 g'_1 + 5 \widetilde{g})\, S_W\, + e\, S'\, (3 - 8\, S_W^2)}{12\, C_W\, S_W}\, ,\\\label{up-CA}
C^u_A(g'_1,\widetilde{g}) &=& -\frac{e\, S'\, + C'\, C_W\, \widetilde{g}\, S_W}{4\, C_W\, S_W}\, ,\\\label{up-width}
\Gamma _u (g'_1, \widetilde{g}) \equiv \Gamma (Z'\to u\overline{u})  &\sim& \frac{3 M_{Z'}}{12 \pi} \left( \frac{1}{144} \left(4g'_1 + 5\widetilde{g} \right)^2 + \frac{1}{16}\widetilde{g}^2\right)\, .
\end{eqnarray}

{\bf Down-type quarks ($d$):}
\begin{eqnarray}\label{down-CV}
C^d_V(g'_1,\widetilde{g}) &=& \frac{C'\, C_W\, (-4 g'_1 + \widetilde{g})\, S_W + e\, S'\, (-3 + 4\, S_W^2)}{12\, C_W\, S_W}\, ,\\\label{down-CA}
C^d_A(g'_1,\widetilde{g}) &=& \frac{C'\, \widetilde{g}\, C_W\, S_W + e\, S'}{4\, C_W\, S_W}\, ,\\\label{down-width}
\Gamma _d (g'_1, \widetilde{g}) \equiv \Gamma (Z'\to d\overline{d})  &\sim& \frac{3 M_{Z'}}{12 \pi} \left( \frac{1}{144} \left(4g'_1 - \widetilde{g} \right)^2 + \frac{1}{16}\widetilde{g}^2\right)\, .
\end{eqnarray}

{\bf Charged leptons ($\ell$):}
\begin{eqnarray}\label{lep-CV}
C^\ell_V(g'_1,\widetilde{g}) &=& \frac{-C'\, C_W\, (4 g'_1 + 3 \widetilde{g})\, S_W + e\, S'\, (1 - 4\,  S_W^2)}{4\, C_W\, S_W}\, ,\\\label{lep-CA}
C^\ell_A(g'_1,\widetilde{g}) &=& -\frac{e\, S' + C'\, C_W\, \widetilde{g}\, S_W}{4\, C_W\, S_W}\, ,\\\label{lep-width}
\Gamma _\ell (g'_1, \widetilde{g}) \equiv \Gamma (Z'\to \ell^+\ell^-)  &\sim& \frac{M_{Z'}}{12 \pi} \left( \frac{1}{16} \left(4g'_1 + 3\widetilde{g} \right)^2 + \frac{1}{16}\widetilde{g}^2\right)\, .
\end{eqnarray}
In the previous formulas, $S'(C')=\sin{\theta'}(\cos{\theta'})$ and $S_W(C_W)$ is the sine(cosine) of the weak mixing angle. Due to the smallness of $\theta'$, as required from LEP constraints~\cite{Abreu:1994ria}, $\sin{\theta'}\ll 1$ and $\cos{\theta'}\simeq 1$ have been adopted in the simplified evaluation of the partial widths.

For the evaluation of the total width, we need to include also the light and heavy neutrinos. Their partial widths read as, respectively, 
\begin{eqnarray}\label{neuL-width}
\Gamma _{\nu _l} (g'_1, \widetilde{g}) \equiv \Gamma (Z'\to \nu_l \nu_l)  &\sim& \frac{M_{Z'}}{24 \pi} \frac{1}{4} \left(2g'_1 + \widetilde{g} \right)^2\, ,\\\label{neuH-width}
\Gamma _{\nu _h} (g'_1) \equiv \Gamma (Z'\to \nu_h \nu_h)  &\sim& \frac{M_{Z'}}{24 \pi} (g'_1)^2\, .
\end{eqnarray}
The extra overall factor $1/2$ in the previous formulas reflect the Majorana nature of the neutrinos, that have only axial couplings. The total width is then given by the sum of the partial contributions
\begin{equation}\label{Zp_width}
\Gamma _{Z'} = 3 \left(\Gamma _u +\Gamma _d +\Gamma _\ell +\Gamma _{\nu _l}+ \Gamma _{\nu _h} \right)\, ,
\end{equation}
where all fermions have been taken massless in comparison to the $\mathcal{O}$(TeV) mass for the $Z'$ boson\footnote{In the case of decoupled heavy neutrinos one of course has $\Gamma_{\nu_h}=0$.}.

\subsection{Benchmark models}\label{sec:benchmarks}

From the previous equations, benchmark models can be selected for specific features. Commonly employed in the literature are the `pure' $B-L$ model, that is defined by the condition $\widetilde{g} = 0$ (implying no mixing at the tree level between the $B-L$ $Z'$, sometimes denoted as
$Z'_{B-L}$, and the SM $Z$, also denoted as $Z_{\rm SM}$); the $U(1)_R$ model, for which Left-Handed
(LH) fermion charges vanish (recovered here by the condition $\widetilde{g}=-2g'_1$); the `$SO(10)$-inspired' $U(1)_\chi$ model, that in our notation reads $\widetilde{g}=-\frac{4}{5}\, g_1'$ (the only orthogonal $U(1)$ extension of the SM hypercharge).

The smallness of $\theta'$ allows us to neglect terms proportional to $\sin{\theta'}$ in eqs.~(\ref{up-CV})--(\ref{lep-CA}), which elucidates  the qualitative features of these common benchmark models. The $\chi$ model will have $C^u_V\sim 0$; the $R$ model will have equal couplings (in absolute value) to all the fermions ($|C^u_V|\sim |C^u_A|\sim |C^d_V|\sim |C^d_A|\sim |C^\ell_V|\sim |C^\ell_A|$) while
the pure $B-L$ model (for which $\widetilde{g}=\sin{\theta'}=0$) will have $C^u_A=C^d_A=C^\ell_A\equiv 0$.

Spatial asymmetries have the following features: they vanish at the $Z'$ peak if any of the chiral couplings are zero and  are very small around it mainly due to the interference with the SM background. In our framework, the axial couplings are proportional to $\widetilde{g}$, and hence vanish identically only in the pure $B-L$ model. Regarding the vector couplings, benchmark scenarios can now be identified by requiring some of them to vanish. Beside the $U(1)_\chi$ model, already known in the literature, the $\notB$ (``B-not'') and the $\notL$ (``L-not'') models are here defined\footnote{Not to be confused with the models in which the baryon number and the lepton number are separately gauged.}. These three models are characterised by the vanishing vectorial coupling to the up-type quarks, 
down-type quarks and charged leptons,  respectively, and are defined by the following relations:
\begin{eqnarray}
U(1)_\chi \mbox{ model:} \qquad \widetilde{g}=-\frac{4}{5}\, g'_1 &\rightarrow& C^u_V \sim 0\, , \\
\notB \mbox{ model:} \qquad \widetilde{g}=4\, g'_1 &\rightarrow& C^d_V \sim 0\, , \\
\notL \mbox{ model:} \qquad \widetilde{g}=-\frac{4}{3}\, g'_1 &\rightarrow& C^\ell_V \sim 0\, .
\end{eqnarray}
The last two models are newly proposed here. These models will have trivial asymmetries in the corresponding final state (the one with negligible vector coupling) and non-vanishing ones in the other final states. Only the pure $B-L$ model will have trivial asymmetries in all final states. Finally, the $R$ model will have almost the same values for the asymmetries in all final states.

On the scenario line characterised by a specific relation between the gauge couplings, we select specific values for the latter to recover well known benchmark models, some of which correspond to those described in Ref.~\cite{Accomando:2010fz}. It is important to note here that only models that are anomaly-free with the SM particle content plus Right-Handed (RH) neutrinos are included in our framework. Finally, to recover the $SO(10)$-inspired $U(1)_\chi$ point, its gauge couplings have been rescaled by the usual factor to account for gauge coupling unification ($\sqrt{3/8}$ in our notation).

\subsection{Calculation and variables}\label{subsec:calc}
The code exploited for our study of the asymmetries is based on helicity amplitudes, defined through the HELAS
subroutines~\cite{HELAS}, and built up by means of MadGraph~\cite{MadGraph}. Initial state quarks have been taken
as massless while for the (anti)top state we have taken $m_t=172.9$ as pole mass. For the $b$-(anti)quark we have taken
$m_b=4.95$ GeV (again, as pole mass). The electron and muons were taken as massless while for the tauon we have used $m_\tau=1.77$ GeV. 
The Parton Distribution Functions (PDFs) exploited
were CTEQ6L1~\cite{cteq}, with factorisation/renormalisation
scale set to $Q=\mu=M_{Z'}$. VEGAS~\cite{VEGAS} was used for the multi-dimensional numerical integrations.
A separate program was also used for part of the analysis, based on CalcHEP~\cite{calchep}, wherein the model has been independently implemented via the LanHEP module~\cite{lanhep} and the FeynRules~\cite{feynrules} package. The model files can be found on the HEPMDB database~\cite{HEPMDB} and on the FeynRules website~\cite{Feynrules_database}.

For all final states, $f=\ell,Q$, a cut in the invariant mass window around the $Z'$ peak was performed
\begin{equation}\label{cut:peak}
\left| m(f\overline{f})-M_{Z'} \right| < 100 \mbox{ GeV}\, ,
\end{equation}
which enhances the $Z'$ peak with respect to the SM background. Regarding the $b$-quark final state, this is still not sufficient to isolate the resonance. Consequently, a further selection has been implemented:
\begin{equation}\label{cut:bb}
p_T(b) > 300 \mbox{ GeV}\, .
\end{equation}

The charge/spin variables that we are going to study have been described in Ref.~\cite{Basso:2012sz} and we summarise here their salient features. The dependence on the chiral couplings of the asymmetries can be expressed analytically, using helicity formulae from Ref.~\cite{Arai:2008qa} (also derived independently with the guidance of~\cite{Hagiwara:1985yu}), for a neutral gauge boson exchanged in the $s$-channel.

\subsubsection{Charge asymmetry}\label{subsubsec:charge}
Charge or spatial asymmetry is a measure of the symmetry of a process under charge conjugation. Due to the {\it CP} invariance of the neutral current interactions, this translates into an angular asymmetry at the matrix element level. It can only be generated from the $q\bar{q}$ initial state due to the symmetry of the di-gluon system and it occurs dominantly from NLO QCD interferences, as described in detail in~\cite{QCDasymmetry} in the case of final state quarks, while for leptons the contribution comes only from the EW sector.

The symmetric $pp$ initial state at the LHC necessitates a more suitable definition of such an observable compared to the well-known top quark forward-backward asymmetry employed at the Tevatron. Several possibilities exist, though
as investigated in~\cite{Basso:2012sz}, the spatial asymmetry that delivers the higher sensitivity is the rapidity dependent forward-backward asymmetry, $A_{RFB}$. It uses the rapidity difference of the final state fermion pair, $\Delta y=|y_{f}|-|y_{\bar{f}}|$, and enhances the $q\bar{q}$ initial state parton luminosity via a cut on the rapidity of the fermion--anti-fermion system, $y_{f\bar{f}}$,
\begin{eqnarray}\label{eqn:asy_ARFB}
    A_{RFB}&=&\frac{N(\Delta y > 0)-N(\Delta y < 0)}{N(\Delta y > 0)+N(\Delta y < 0)}\Bigg |_{|y_{f\bar{f}}|>y^{cut}_{f\bar{f}}}\, \\ 
           &\propto& C^i_A C^i_V C^f_A C^f_V \, .
\end{eqnarray}
{In this paper, we used $y^{cut}_{f\bar{f}}=0.5$.}
It is clear that this observable can only be generated by a $Z'$ boson if its vector ($C_V$) and axial ($C_A$) couplings to both the initial ($i$) and final ($f$) state fermions are non vanishing. We will study $A_{RFB}$ for light leptons ($\ell = e,\,\mu$) and for the heavy quark final states.

\subsubsection{Spin asymmetries}\label{subsubsec:spin}
Spin asymmetries, on the other hand, focus on the helicity structure of the final state fermions and, when such properties are measurable, display interesting dependences on the chiral structure of the $Z^\prime$ boson couplings. The helicity of a final state can only be experimentally determined for a decaying final state, where the asymmetries are extracted as coefficients in the angular distribution of its decay products. This is described for the case of top quarks in Ref.~\cite{tt-pol}, but can be easily extended to other weakly decaying particles.

We define two such asymmetries. The main observable we consider is the polarisation or single spin asymmetry, $A_L$, defined as follows:
\begin{eqnarray}\label{eqn:asy_AL}
    A_{L}&=&\frac{N(-,-) + N(-,+) - N(+,+) - N(+,-)}{N_{Total}}\, ,\\
         &\propto& C^f_A\, C^f_V \,\beta\,\Big( (C^i_V)^2 + (C^i_A)^2 \Big)\, ,
\end{eqnarray}
where $\beta=\sqrt{1-4\, m_f^2/\hat{s}}$ and $N$ denotes the number of observed events and its first(second) argument corresponds to the helicity of the final state particle(anti-particle). It singles out one final state particle, comparing the number of positive and negative helicities, while summing over the helicity of the other anti-particle (or vice versa).
This observable is proportional to the product of the vector and axial couplings of the final state only and is therefore additionally sensitive to their relative sign, a unique feature among asymmetries and cross section observables. { This is equivalent to a sensitivity to the relative right- or left-handedness of the $Z'$ couplings. As described in the introduction, it can be measured now for $\tau$ and $t$-quark final states while for $b$-quarks an improved vertex detector would probably be needed, so that we only present the latter case here as a suggested analysis.}

In the case of appreciably massive final states, i.e., the top quark only, the spin correlation or double spin asymmetry is accessible. This observable relies on the helicity flipping of one of the final state particles, whose amplitude is proportional to $m_{f}/\sqrt{\hat{s}}$, where $\sqrt{\hat{s}}$ is the partonic centre-of-mass (CM) energy, and gives the proportion of like-sign final states against the opposite-sign states,
\begin{eqnarray}\label{eqn:asy_ALL}
    A_{LL}&=&\frac{N(+,+) + N(-,-) - N(+,-) - N(-,+)}{N_{Total}}\, ,\\
    &\propto& \Big( 3\, (C^f_A)^2\beta^2 + (C^f_V)^2(2+\beta^2)\Big) \, \Big( (C^i_V)^2 + (C^i_A)^2 \Big) \, .
\end{eqnarray} 
This observable depends only on the square of the couplings in a similar way to the total cross section. In the massless limit, $\beta\rightarrow 1$ and $A_{LL}$ becomes maximal, making it a relevant quantity to measure only in the $t\bar{t}$ final state. 

\subsubsection{Asymmetries and $Z'$ couplings}
{ 
The ultimate scope of profiling a $Z'$ boson is that of measuring its couplings to fermions. Although it is beyond the scope of this paper, we still would like to make some comments. In the minimal case considered here, $5$ independent parameters exist ($q_L$, $u_R$, $d_R$, $\ell_L$, $e_R$, where here $q_L$ and $\ell_L$ identify the quark and lepton doublets, respectively), so that $5$ independent measurements are required. In 
 Ref.~\cite{Petriello:2008zr}, it is shown that a set of $4$ coefficients ($c_q$, $e_q$; $q=u,d$) that are functions of the $Z'$ couplings can be extracted from 4 observables related to charge asymmetry and total cross section in the light lepton channel (electrons and muons), by restricting the kinematical domain of the differential cross section. It is observed that a degeneracy exists between the dependence on leptonic and quark couplings in this analysis, stemming from the minimal assumption of 5 independent parameters and only 4 independent observables.
Including the total $Z'$ width does not seem to fix the problem, since it enlarges the set of independent parameters to include all the fermions, such as the SM light neutrinos and all the coupled exotic particles, in our case the heavy neutrinos. Instead, spin observables, being based on polarised amplitudes, cannot be obtained from the fully differential cross section and hence can be considered. Staying on the leptonic final state, the tau polarisation is the first case to look at.
It has a different dependence on the quark and lepton couplings than $c_q$, $e_q$. Recasting eq.~\eqref{eqn:asy_AL}, it can be written in the spirit of~\cite{Petriello:2008zr} as a new coefficient,
\begin{align}
      f_{q=u,d}\propto(q^{2}_{R}+q^{2}_{L})(e^{2}_{R}-e^{2}_{L}),
\end{align}
making manifest the aforementioned sensitivity to the `handedness' of the $Z'$ couplings. These coefficients can be extracted from tau polarisation measurements in a similar way as the $c_q$ and $e_q$ coefficients are from the differential cross section, i.e., by splitting the kinematical domain of $A_L^\tau$  in 2 independent regions to then be fitted. It is evident that the same quark-lepton degeneracy exists and in fact $ f_{q=u,d}$ turns out not to be linearly independent from the $c$ and $e$ coefficients when $u_{L}=d_{L}$:
\begin{align}
      f_{q=u,d}\propto c_{q}\frac{e_{u}-e_{d}}{c_{u}-c_{d}}.
\end{align}
Although the tau polarisation does not allow for the extraction of the $Z'$ couplings within the 5 parameter minimal assumption, it still proves to be a useful quantity to measure. First of all, it can be used as a test of universality, having better sensitivity than the charge asymmetries in a tau final state. Secondly, it provides an extra set of constraints on the other coefficients in order to improve the quality of a fit of the couplings. The ratios
\begin{align}
      \frac{f_{u}}{c_{u}}=\frac{f_{d}}{c_{d}}\equiv \frac{e^{2}_{R}-e^{2}_{L}}{e^{2}_{R}+e^{2}_{L}}
\end{align}
should be equal and are independent of the quark couplings and could serve to reduce some systematics, such as the one originating from the PDFs. 
To break the degeneracy allowing for a direct fit of the couplings, one must then consider observables in non-leptonic final states, \emph{e.g.}, top polarisation or charge asymmetry. For the longer term, $b$-quark final states may turn out to be useful too.
Moving away from the minimal assumption permitting $u_{L}\neq d_{L}$ (but still requiring universality), as could occur with $Z'$s arising from more general gauge group extensions, restores the linear independence of one of the tau polarisation coefficients. In this case one still requires the use of an alternative final state to complete the set of independent observables needed to fit directly to the $6$ (or more) couplings, to that the exploitation of top (and possibly bottom) final states is then called for again.}

\subsubsection{Efficiencies and uncertainties}\label{subsubsec:effs}
To be able to quantitatively address the distinguishability among the various models and the SM background, we associate a statistical error to each observable.
Although in this work we only estimate statistical uncertainties,
systematics may also be important~\cite{Hewett:2011wz,Alvarez:2012vq} though the mass window selection is expected to milden their actual contribution. However, their inclusion would require detailed detector simulations which are beyond the scope of this paper. In this respect, it should further be noted that, by the time the LHC reaches the 14 TeV stage, where our most interesting results are applicable, systematics will be much better understood than at present.

 Given that an asymmetry is defined in terms of the number of events measured in
some generic `forward' ($N_{F}$) and `backward' ($N_{B}$) directions (this is also true for spin asymmetries), the statistical error is evaluated by 
propagating the Poisson error on each measured 
quantity (i.e., $\delta N_{F(B)}=\sqrt{N_{F(B)}}$). Per given integrated luminosity $\mathcal{L}$, the measured number of events will be $N=\varepsilon\mathcal{L}\sigma$, $\sigma$ being the cross section, yielding an uncertainty on the asymmetry $A$ of 
\begin{equation}\label{eqn:error}
    \delta A\equiv \delta\left(\frac{N_{F}-N_{B}}{N_{F}+N_{B}}\right)=\sqrt{\frac{2}{\mathcal{L}\varepsilon}    
\left(\frac{\sigma^{2}_{F}+\sigma^{2}_{B}}{\sigma_{Total}^{3}}\right)}.
\end{equation}
 Here, $\varepsilon$  corresponds to the reconstruction efficiency of the final state system. For the $t\overline{t}$ system, we take $\varepsilon_t=10\%$ as in \cite{Basso:2012sz}, considering all possible decay channels, based on efficiencies quoted by recent experimental papers~\cite{top-eff} as well as estimates from Monte Carlo studies in previous works~\cite{Frank:2011rb}. For the $b\overline{b}$ system, we take $\varepsilon_b=5\%$. 
 { This number is based on an LHC $b$-tagging efficiency of $50\%$, given by summing 1-tag and 2-tags efficiencies at high $p_T$~\cite{CMS-11-008}, to extract the $b$-quark decay mode from the di-jet sample. The latter reference shows that, while the 2-tags do not yield measurable rates, the 1-tag alone seems insufficient, given the almost equal rate of mistags. However, it is shown in~\cite{CMS-11-004} that additional requirements on a muon from the $b$-jet is capable of improving both the tagging and the tagging-over-mistag efficiencies for $b$-quarks up to intermediate jet $p_{T}$ while no information is available for very energetic jets. Regarding the $b$-hadron helicity, there is no result currently available at the LHC, thus we take the values available from LEP~\cite{BPOL,LEP-b}, quantifiable in an overall $10\%$ efficiency, as an estimate to quantitatively discuss spin observables. By considering this extra efficiency factor, we are in fact also further selecting $b$-jets with prompt hard leptons. We base our overall reconstruction efficiency estimate on all of these considerations hoping that, for the time scales at which this paper could be relevant, the situation will have improved as suggested by the proposed new generation of micro-vertex detector available for the $\sqrt{s}=14$ TeV run, see, e.g., Ref~\cite{PoS(VERTEX)}. However, we remind the reader that these observables are just a suggested measure and might not be available at the LHC, although the proposed upgrades for micro-vertex detectors may well help to extract them.}
  For the $\ell^+\ell^-$ system, we distinguish between light leptons ($e,\mu$), for which $\varepsilon_{e,\mu}=90\%$~\cite{lep_eff}, and the taus, for which $\varepsilon_{\tau}=10\%$. This latter number is based on a recent ATLAS measurement in 1--prong decays ($25\%$ of tau's Branching Ratios (BRs)), its reconstruction efficiency of $60\%$ and a $20\%$ loss of sensitivity due to detector effects~\cite{TAULHC}. We remark here again that light leptons will be employed for charge asymmetries only, whilst we will make use of taus to define the spin asymmetry $A_L$ in the leptonic final state.

All figures that will be shown in the following are for the LHC at $\sqrt{s}=14$ TeV assuming 100 fb$^{-1}$ of integrated luminosity, for a $Z^\prime$ mass of 2.5 TeV and decoupled heavy neutrinos. In some cases, observables are evaluated also for `light' heavy neutrinos of $50$ GeV mass to estimate their impact. Values of an observable in the ($g'_1,\widetilde{g}$) coupling plane can be compared to SM predictions corresponding to the point $(0,0)$. For the signal, the uncertainty for that value is considered to confirm visibility over the background in that channel. Also shown are surface plots of the significance of the observable with respect to the SM prediction, defined as
\begin{equation}\label{eqn:sig}
s \equiv \frac{\left| A(1)-A(2)\right|}{\sqrt{\delta A(1)^2 + \delta A(2)^2}}\,,
\end{equation}
where $A(1)$ and $A(2)$ denote the prediction for an observable by two different hypotheses and $\delta$ refers to the statistical uncertainty. In this case, the statistical uncertainty of the SM prediction is taken to be zero because its very low cross section at such high invariant masses (especially true for the leptonic final state) would artificially dilute the significance of asymmetries in that region. Therefore, in the hypothesis of observing a $Z'$ boson, we prefer to consider the SM prediction as a reference value only.

%% file: sect3.tex
We then move on to profiling the $Z'$ boson analysing its BRs, cross section and asymmetries in given final states.

\subsection{Exclusion limits}

\begin{figure}[h]
  \begin{center}
  \includegraphics[width=0.7\textwidth,angle=0]{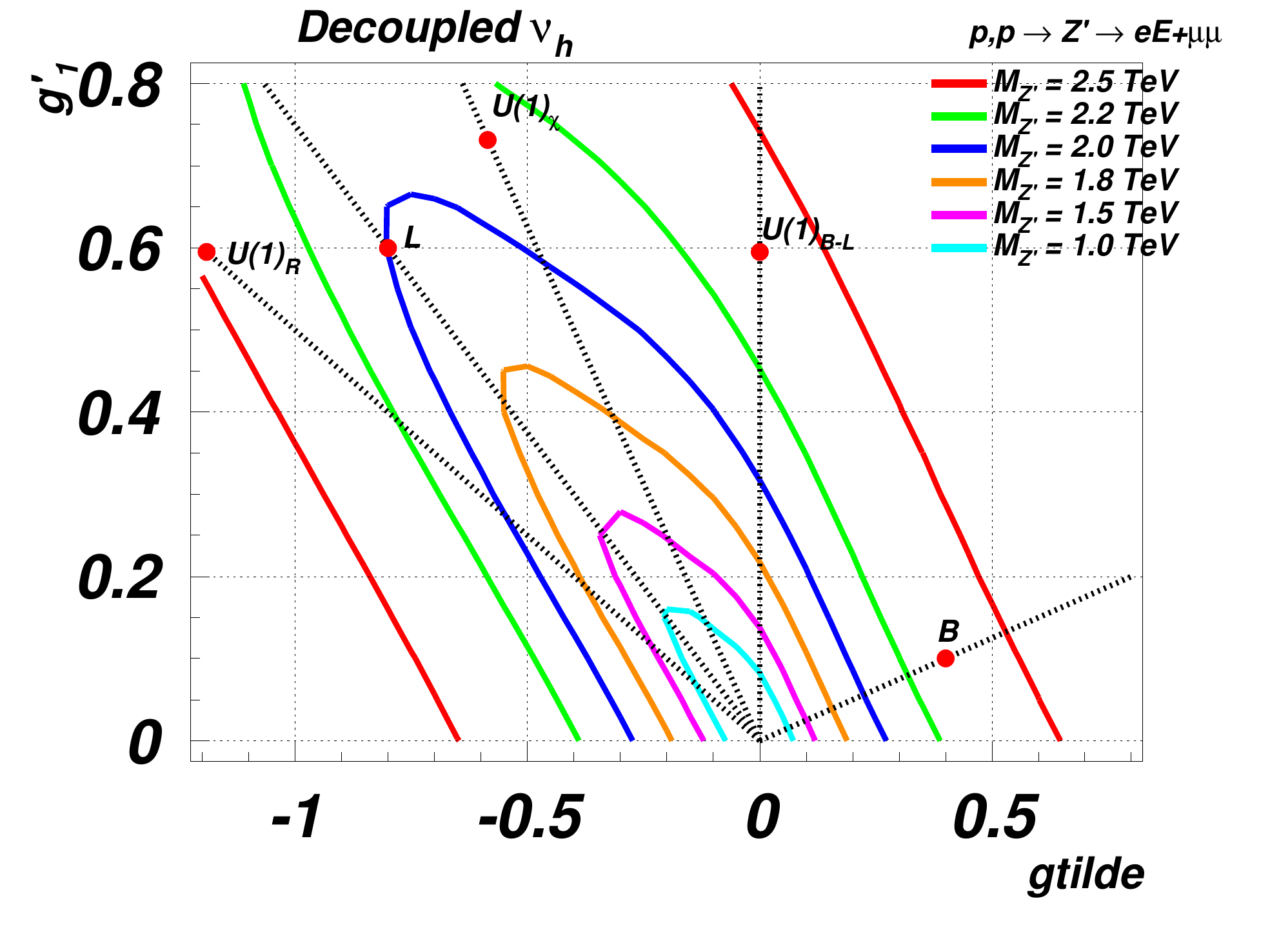}
  \end{center}
  \caption{$Z'$ exclusions from CMS latest data, at $\sqrt{s}=7$ TeV for the combination of $4.7$ fb$^{-1}$ in the electron channel and of $4.9$ fb$
^{-1}$ in the muon channel. The dotted black lines refers to the main benchmark models of this analysis (see section~\ref{sec:benchmarks}).}
\label{Zp-excl}
\end{figure}

To start with, we present (see figure~\ref{Zp-excl}) the most recent $95\%$ (Confidence Level) CL exclusions at the LHC in the ($g'_1,\widetilde{g}$) plane (first in~\cite{Basso:2012ti}), based on the CMS data at $\sqrt{s}=7$ TeV for the combination of $4.7(9)$ fb$^{-1}$ in the electron(muon) channels~\cite{LHC}.

Table~\ref{tab:exclusions} collects the maximum allowed $g'_1$ coupling per given $Z'$ boson mass for the various benchmark model of interest (see section~\ref{sec:benchmarks}).

\begin{table}[t!]
        \centering
        \begin{tabular}{|c|c|c|c|c|c|}\hline
$M_{Z'}$ (TeV) & $U(1)_R$ & $U(1)_\chi$ & $U(1)_{B-L}$ & $\notL$ & $\notB$\\
        \hline
2.5 & $0.63$  & $>0.8$ & $0.75$ & $>0.8$ & $0.13$\\
2.2 & $0.39$  & $0.81$ & $0.45$ & $0.83$ & $0.08$\\
2.0 & $0.27$  & $0.58$ & $0.31$ & $0.60$ & $0.06$\\
1.8 & $0.19$  & $0.40$ & $0.22$ & $0.41$ & $0.04$\\
1.5 & $0.12$  & $0.24$ & $0.13$ & $0.25$ & $0.03$\\
1.0 & $0.075$ & $0.14$ & $0.08$ & $0.15$ & $0.02$\\ \hline
        \end{tabular}
\caption{Maximum $g'_1$ allowed at $95\%$ CL for the benchmark lines of 
section~\ref{sec:benchmarks}.\label{tab:exclusions}}
\end{table}

\subsection{BRs, total widths and production cross sections}

\begin{figure}
\centering
 \includegraphics[angle=0,width=0.49\textwidth ]{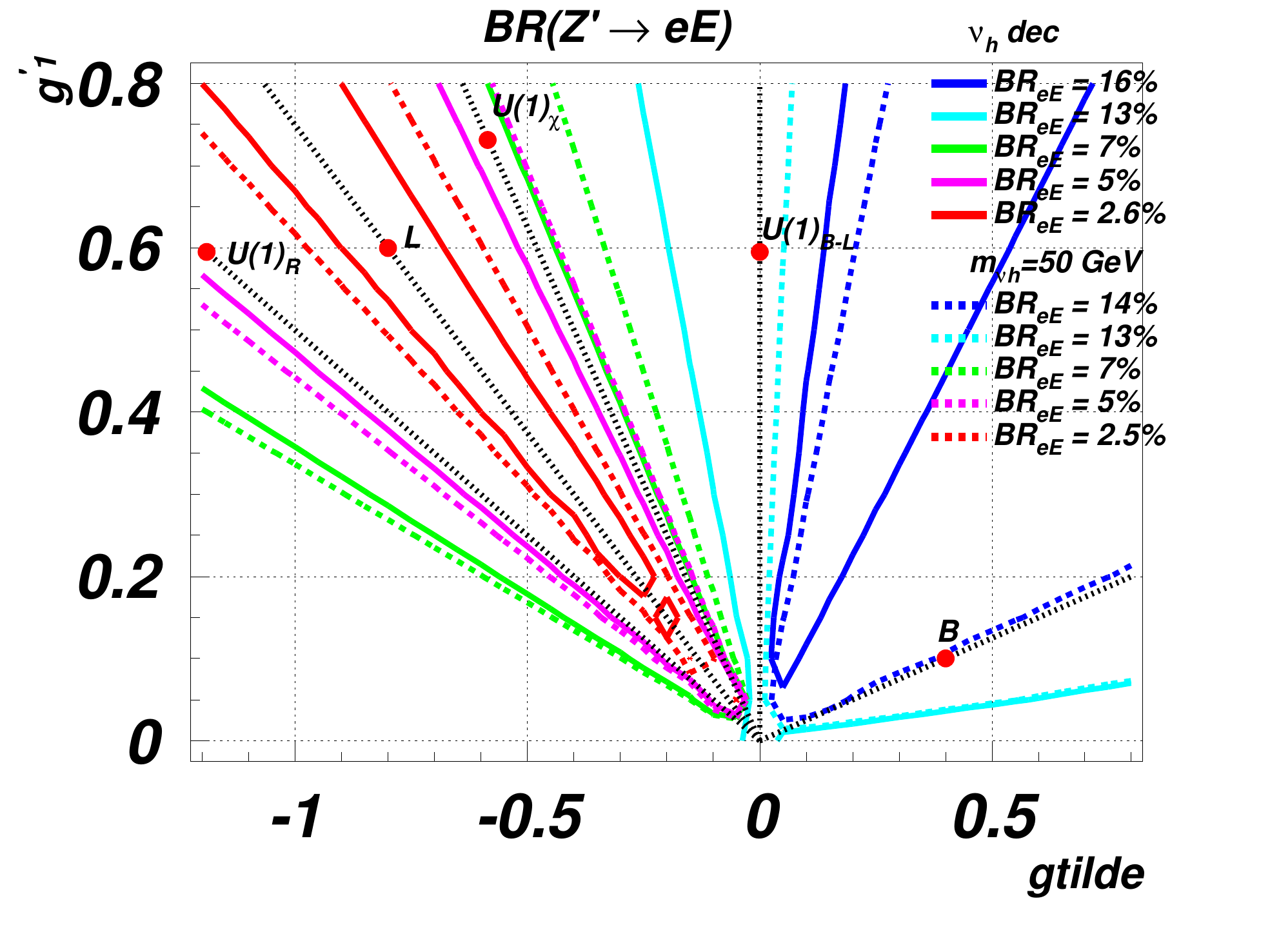} \includegraphics[angle=0,width=0.49\textwidth ]{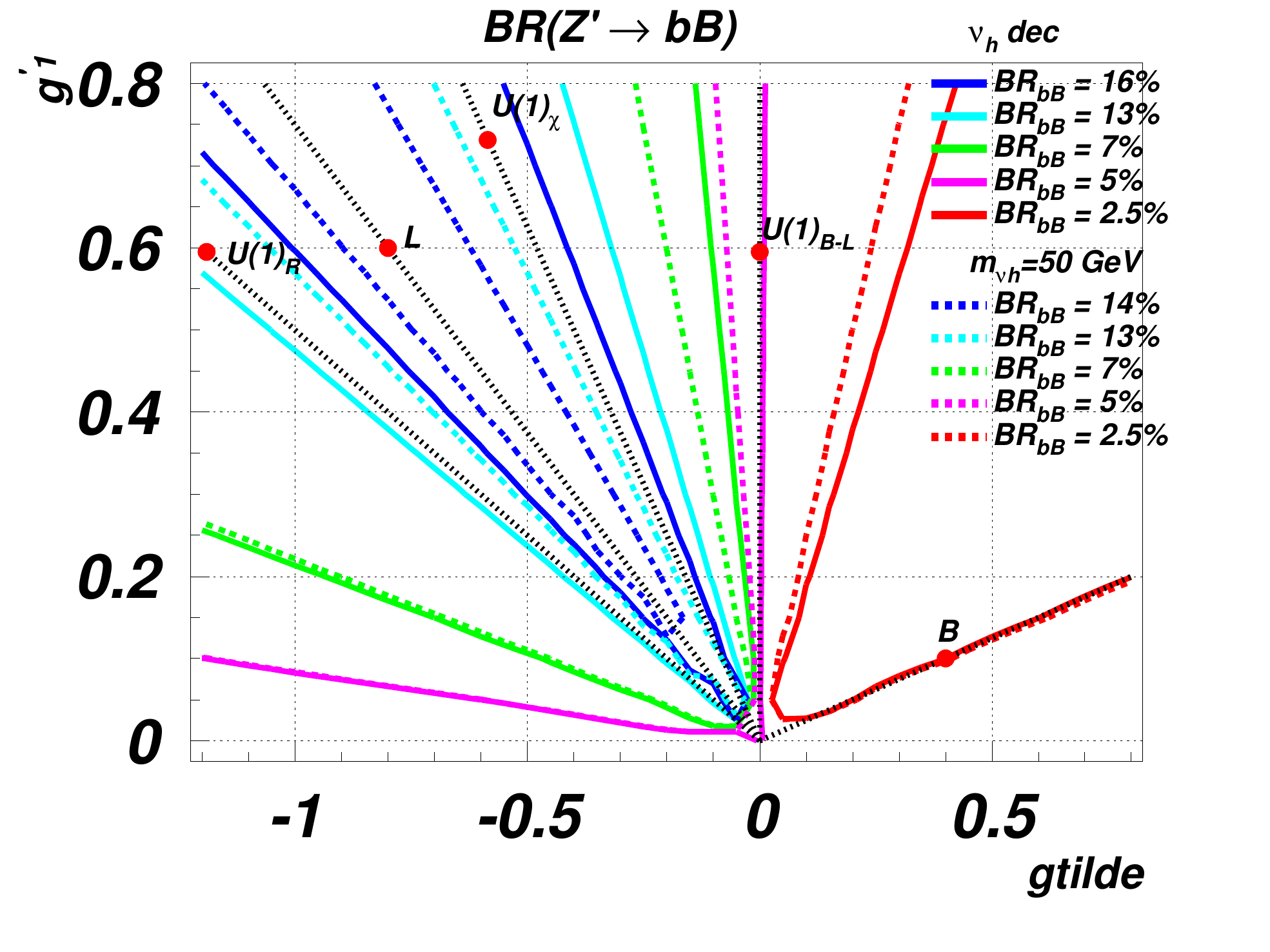} \\
 \includegraphics[angle=0,width=0.49\textwidth ]{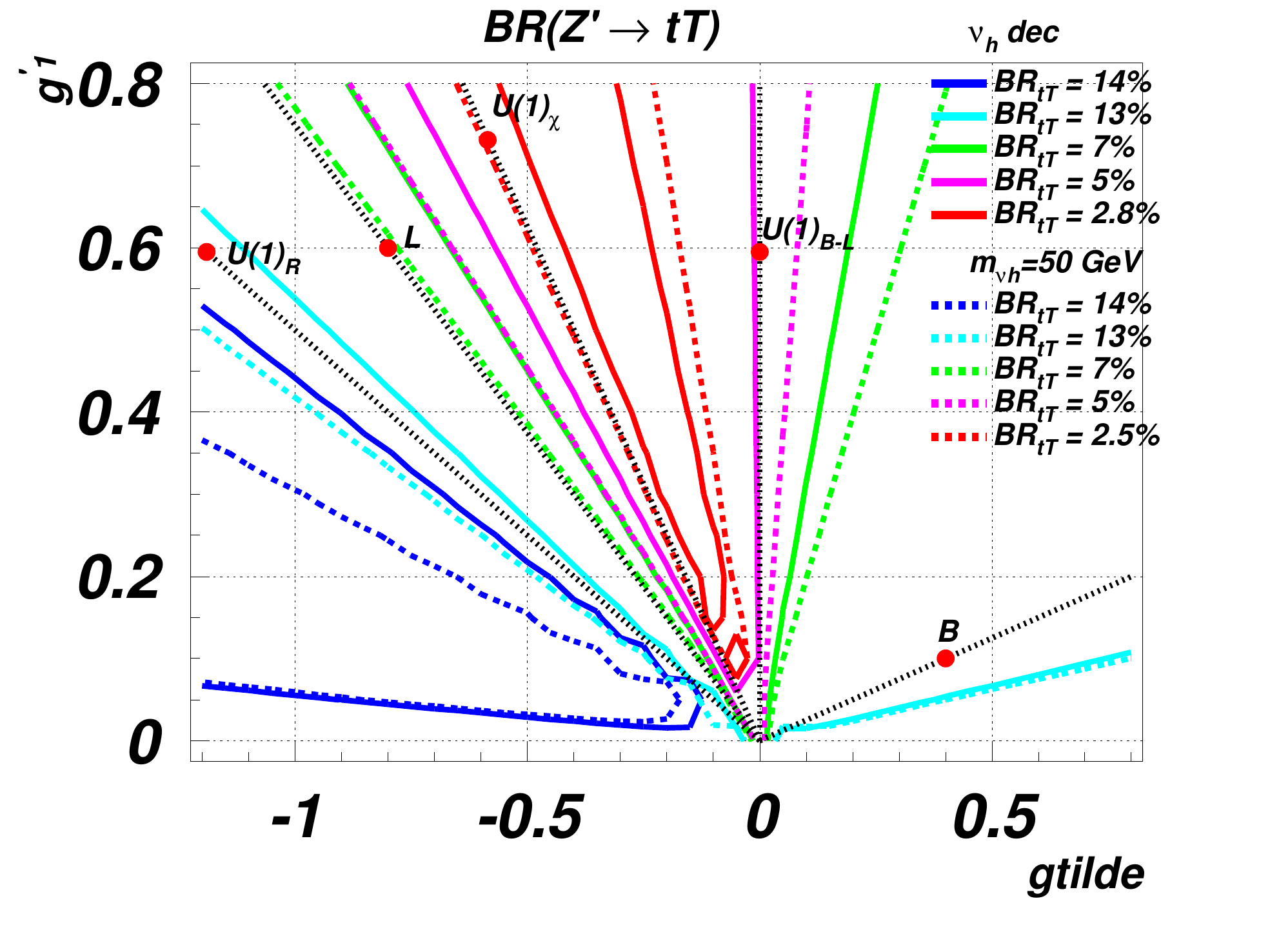}
 \includegraphics[angle=0,width=0.49\textwidth ]{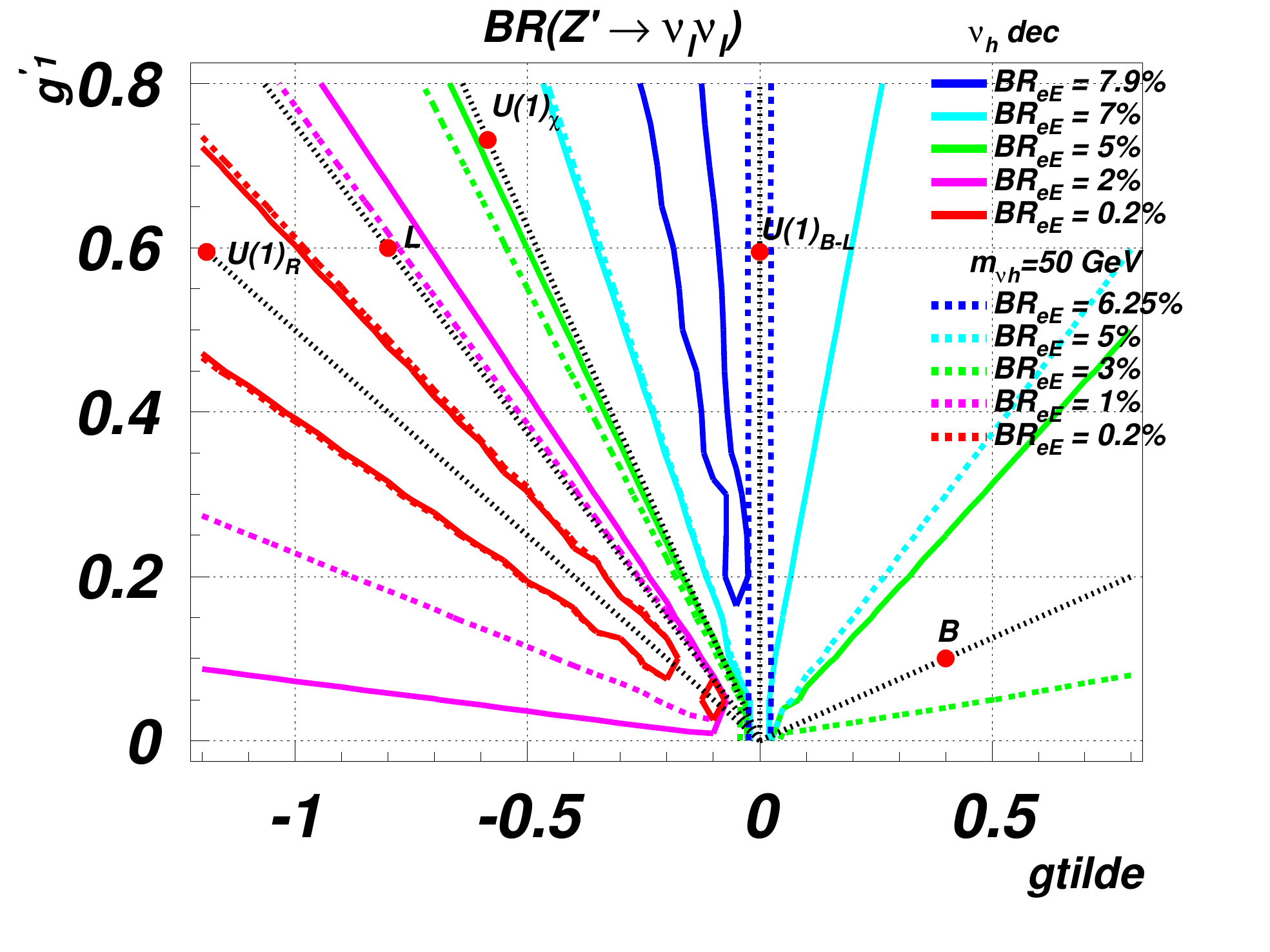}\\
\includegraphics[angle=0,width=0.49\textwidth ]{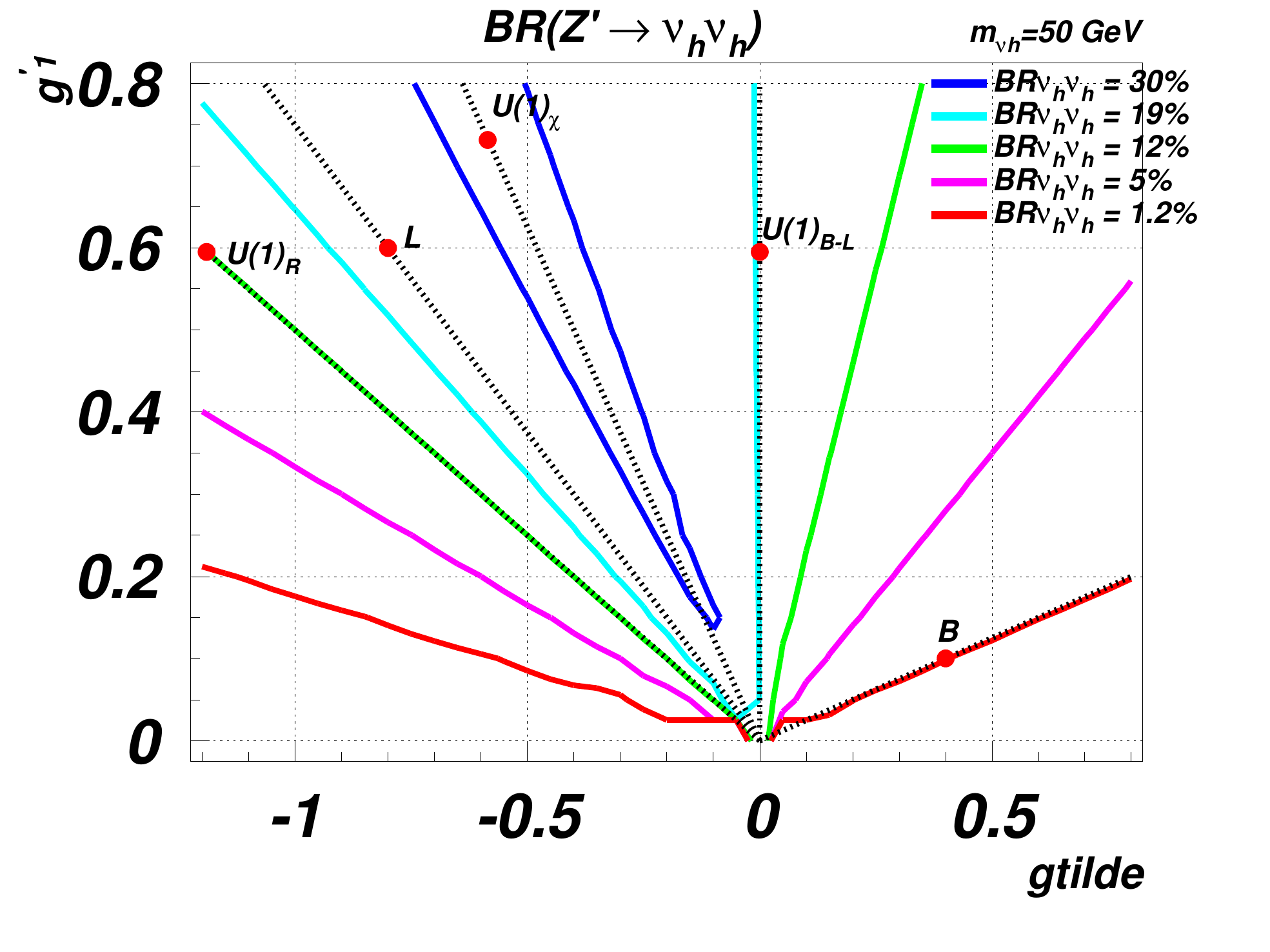}
 \caption{\label{BRs} BRs per each generation (top row) for $Z'\to e^+e^-$ and for $Z'\to b\overline{b}$, (middle row) for $Z'\to t\overline{t}$ and
 for $Z'\to \nu_l \nu_l$, and (bottom row) for $Z'\to \nu_h \nu_h$, in the ($g'_1$, $\widetilde{g}$) plane for two extreme neutrino masses, i.e., for 
decoupled neutrinos and for very light neutrinos ($m_{\nu _h}=50$ GeV). The benchmark models are highlighted as (red) dots on the (dotted) scenario 
lines.\label{fig:BRs}}
\end{figure}

We start by analysing the BRs of the individual final states: up- and down-type quarks, charged leptons and light and heavy neutrinos. Minima and maxima are easily found by minimising the respective BR. We chose to evaluate them as a function of $\widetilde{g}$:
\begin{equation}
\frac{\partial}{\partial\widetilde{g}} \frac{\Gamma _f}{\Gamma _{Z'}}(g'_1,\,\widetilde{g}) \equiv 0\, ,
\end{equation}
so to obtain relations of the type $\widetilde{g}=\widetilde{g}(g'_1)$. The results with maximum and minimum expected BRs are collected in table~\ref{tab:BRs} and are displayed in figure~\ref{fig:BRs}.

\begin{table}[t!]
        \centering
        \begin{tabular}{|c|c|c||c|c|}\hline
Dec. $\nu_h$ & \multicolumn{2}{c||}{maximum} & \multicolumn{2}{c|}{minimum}\\
        & $\widetilde{g}=\widetilde{g}(g'_1)/g'_1$ & BR ($\%$) & $\widetilde{g}=\widetilde{g}(g'_1)/g'_1$ & BR  ($\%$)\\  \hline
$Z'\to u\overline{u} $   & $\frac{1}{24}\left( -47-\sqrt{1153}\right)$ & $15.3$& $\frac{1}{24}\left( -47+\sqrt{1153}\right)$ & $2.5$ \\
$Z'\to d\overline{d}$   & $\frac{1}{8}\left( 1-\sqrt{97}\right)$ & $20.5$& $\frac{1}{8}\left( 1+\sqrt{97}\right)$ & $1.85$ \\
$Z'\to \ell^+\ell^- $& $\frac{1}{40}\left( -15+\sqrt{1345}\right)$ & $16.2$& $\frac{1}{40}\left( -15-\sqrt{1345}\right)$ & $2.3$ \\\
$Z'\to \nu_l\nu_l$& $-\frac{1}{4}$ & $8.0$ & $-2$  & $0$ \\ \hline 
%
\hline
$m_{\nu_h}$ & \multicolumn{2}{c||}{maximum} & \multicolumn{2}{c|}{minimum}\\
$50$ GeV & $\widetilde{g}=\widetilde{g}(g'_1)/g'_1$ & BR ($\%$) & $\widetilde{g}=\widetilde{g}(g'_1)/g'_1$ & BR  ($\%$)\\  \hline
$Z'\to u\overline{u} $   & $\frac{2}{3}\left( -4-\sqrt{10}\right)$ & $15.0$& $\frac{2}{3}\left( -4+\sqrt{10}\right)$ & $1.75$ \\
$Z'\to d\overline{d} $   & $2\sqrt{\frac{2}{5}}$ & $15.0$& $-2\sqrt{\frac{2}{5}}$ & $1.75$ \\
$Z'\to \ell^+\ell^- $& $2\sqrt{\frac{2}{5}}$ & $15.0$& $-2\sqrt{\frac{2}{5}}$ & $1.75$ \\
$Z'\to \nu_l\nu_l$& $0$ & $6.25$ & $-2$  & $0$ \\
$Z'\to \nu_h\nu_h$& $-\frac{4}{5}$ & $10.5$ & $g'_1\to 0$\  & $0$ \\ \hline
        \end{tabular}

\caption{$\widetilde{g}=\widetilde{g}(g'_1)$ that delivers the maximum and minimum BRs and their extreme values (per each generation), for ({\it top
}) decoupled and ({\it bottom}) very light heavy neutrinos.}\label{tab:BRs}
\end{table}

\begin{figure}[!ht]
\centering
 \includegraphics[angle=0,width=0.49\textwidth ]{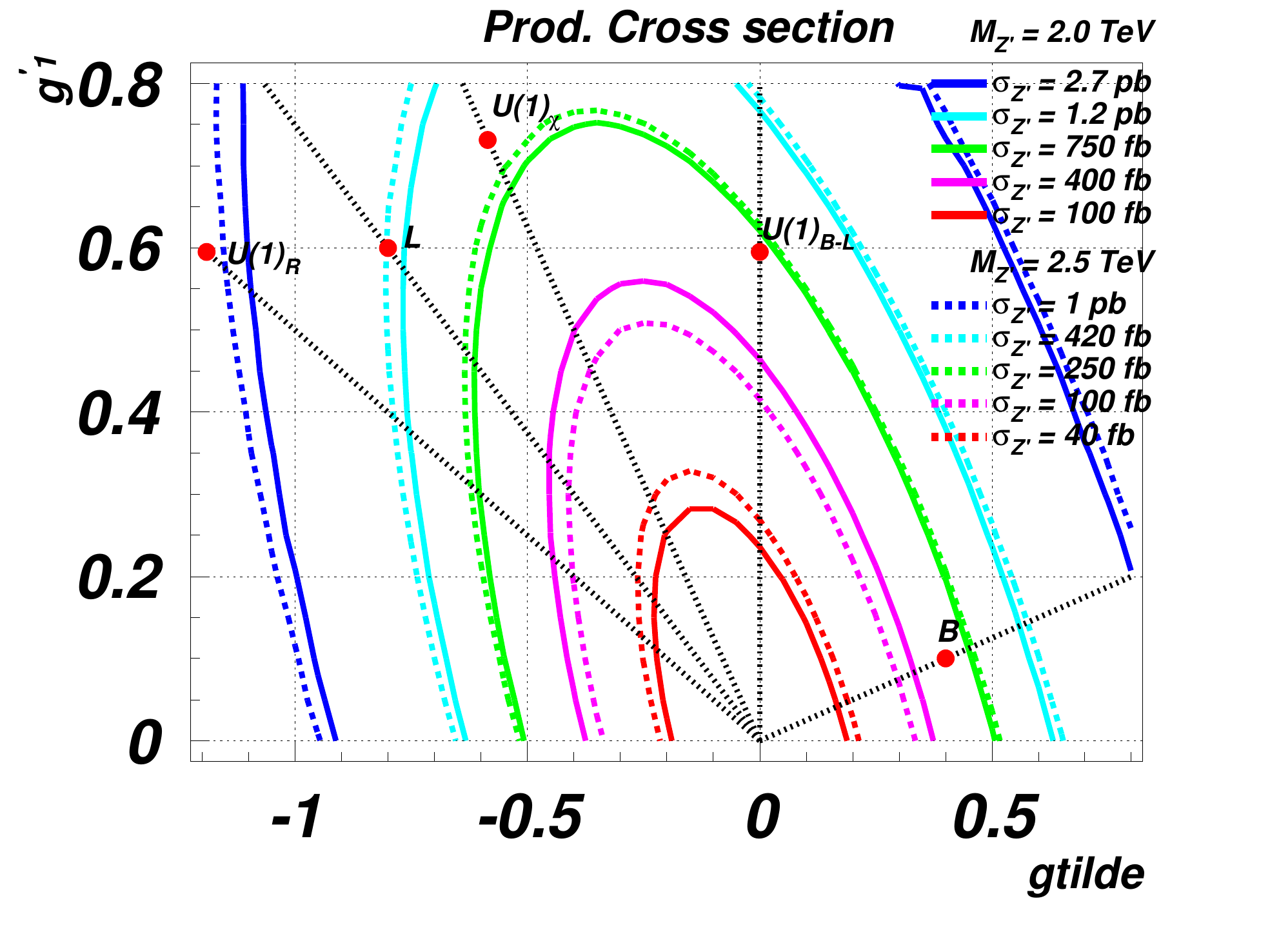}  
 \includegraphics[angle=0,width=0.49\textwidth ]{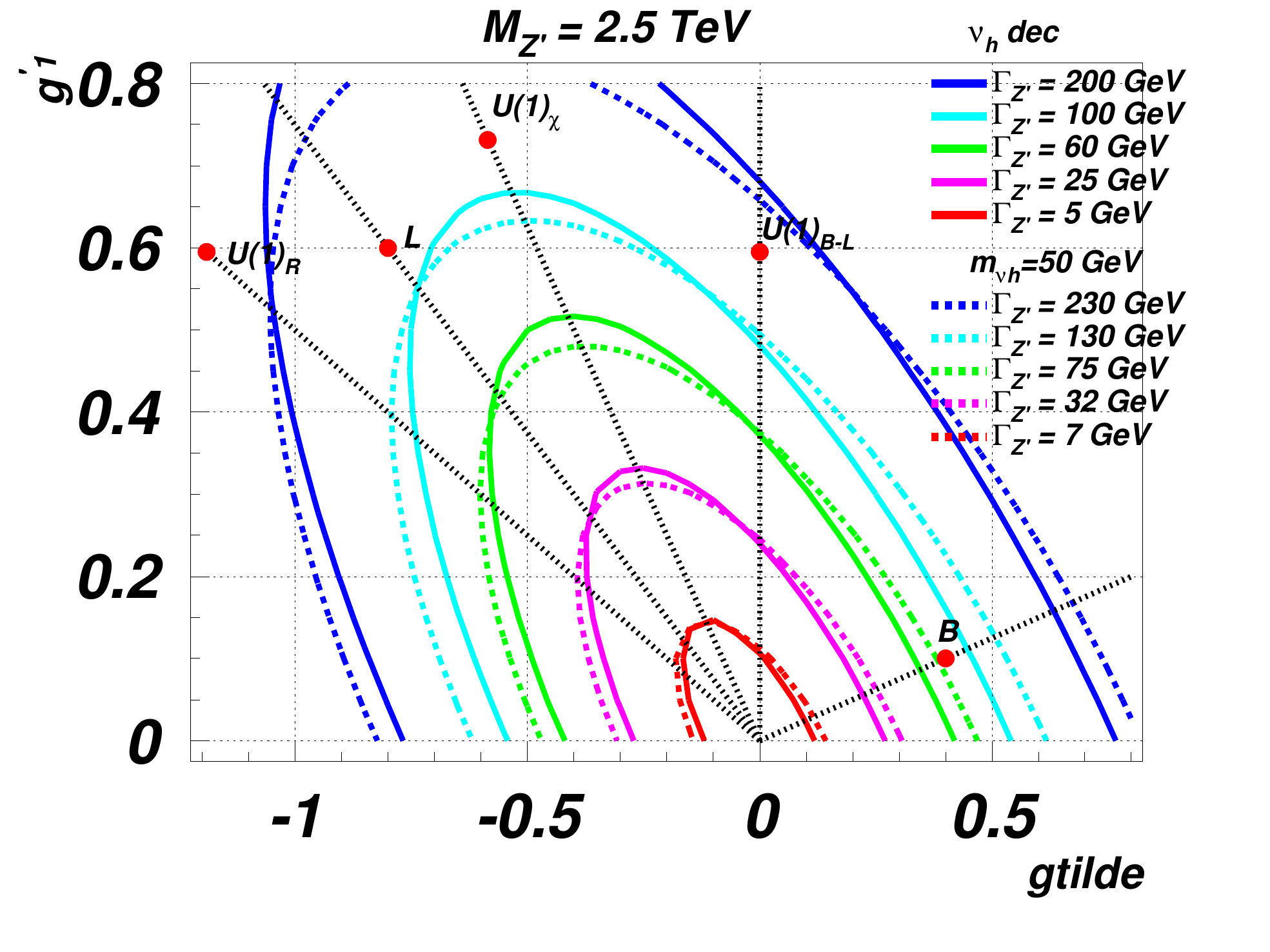} 
 \caption{\label{fig:total_width} Total production cross section (for $\sqrt{s}=14$ TeV) (left) and total width (right) for $M_{Z'}=2.5$ TeV
 in the ($g'_1$, $\widetilde{g}$) plane for two extreme neutrino masses, i.e., for decoupled neutrinos and for very light neutrinos ($m_{\nu _h}=50$ 
GeV). The benchmark models are highlighted as (red) dots on the (dotted) scenario lines. }
\end{figure}

We recognise some features: the $U(1)_R$ scenario has the lowest BR into light neutrinos, regardless of the heavy neutrino masses. This is simply because light neutrinos are predominantly LH fermions, hence they do not couple to the $Z'$ boson in this scenario. When the heavy neutrinos are light, the maximum of the BR($Z'\to \nu_l \nu_l$) is in the pure $B-L$ model. Regarding the heavy neutrinos, they are predominantly RH fields (as already mentioned), hence with zero hypercharge. Consequently, their partial width is independent of $\widetilde{g}$, as in eq.~(\ref{neuH-width}): on the one side, the minimum BR is for $g'_1\to 0$ (when they decouple), on the other side the maximum BR is for the minimum total $Z'$ width. By direct calculation, we find that the total width of eq.~(\ref{Zp_width}) is minimised in the $U(1)_\chi$ scenario 
(figure~\ref{fig:total_width}), as also indicated in table~\ref{tab:BRs} (corresponding to the maximum BR into heavy neutrinos).

Figure~\ref{fig:total_width} (left) shows the production cross sections for two different values of the $Z'$ mass. For large values of the couplings, the production cross section can be up to $\mathcal{O}(1~$pb), in regions of parameter space not excluded experimentally, in accordance with figure~\ref{Zp-excl}. For very small albeit non-vanishing couplings, the production cross section for a $Z'$ boson is easily bigger than tens of fb. These values are some orders of magnitude above the expected SM background when leptons in the final state are considered, as we will see in the following section.

Heavy neutrinos can have a strong impact on the $Z'$ features, altering the lineshape in a measurable way. First of all, the total width is augmented
 by up to a factor $30\%$, when comparing same $Z'$ boson mass and same values for the gauge couplings. A similar factor is also the maximum BR of the $Z'$ boson into heavy neutrinos (once summed over generations).

\subsection{Total event rates}\label{sec:ev_rates}
\begin{figure}[!t]
\centering
\includegraphics[angle=0,width=0.48\textwidth]{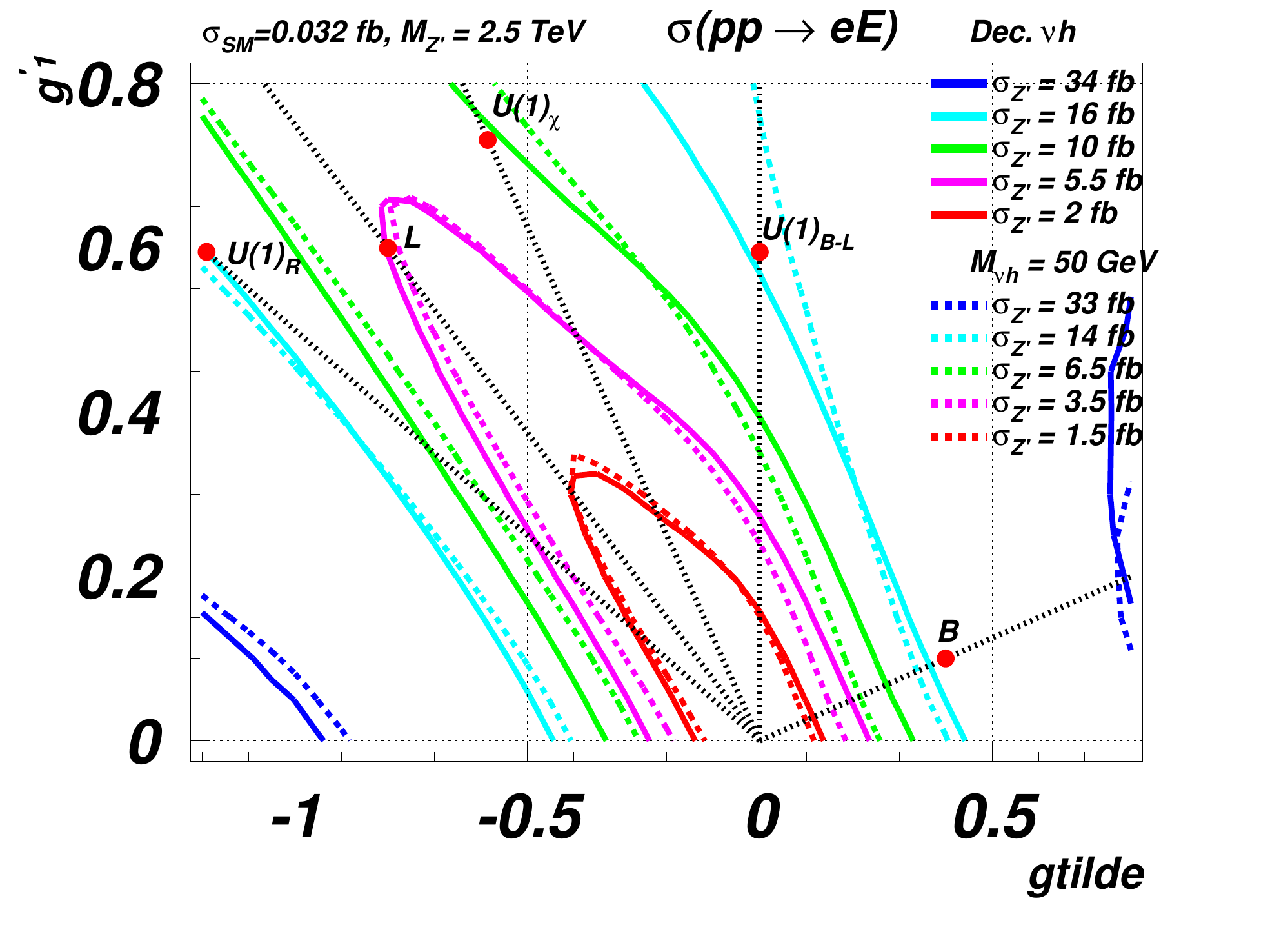}
\includegraphics[angle=0,width=0.48\textwidth]{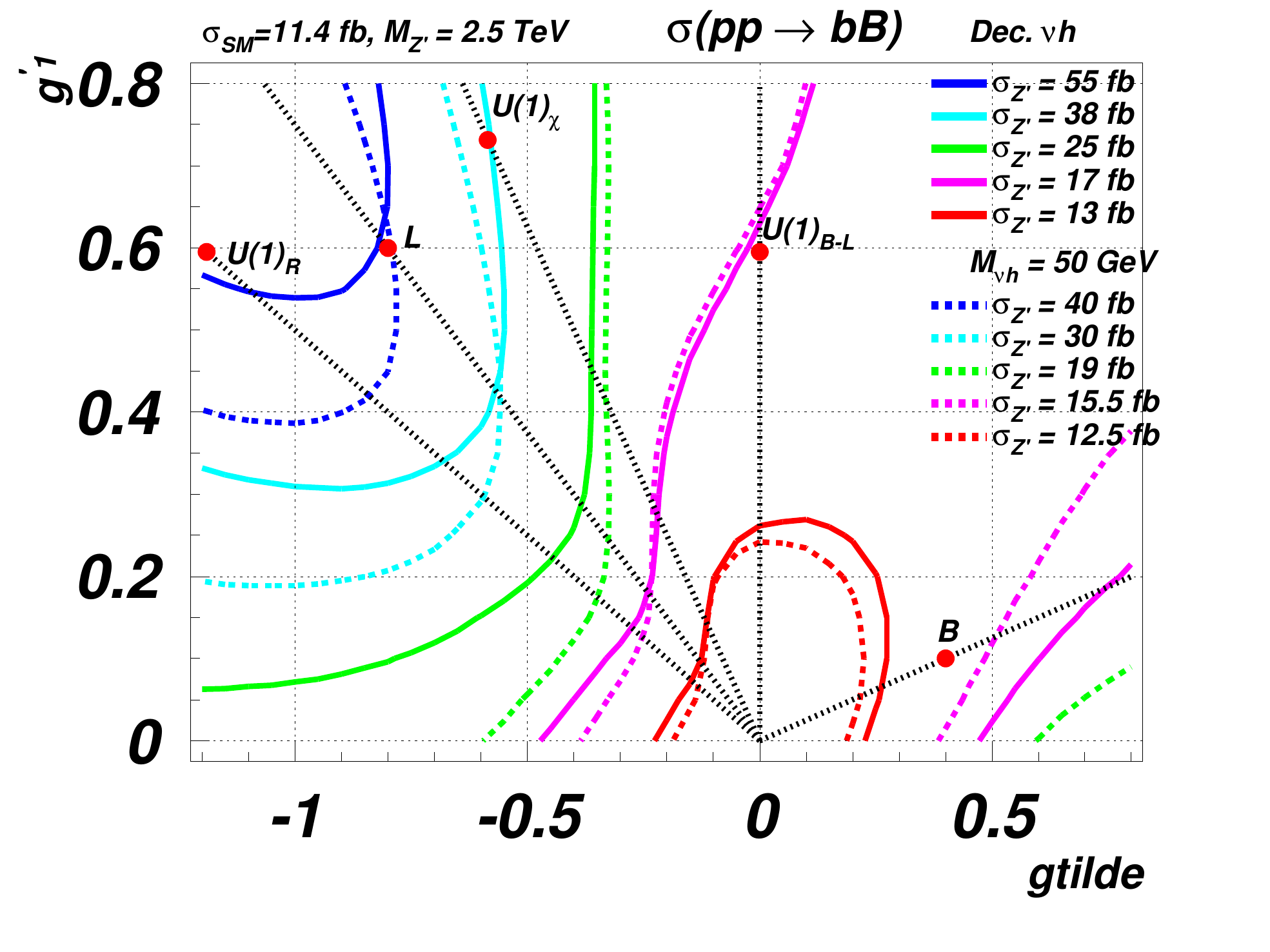}\\
\includegraphics[angle=0,width=0.48\textwidth]{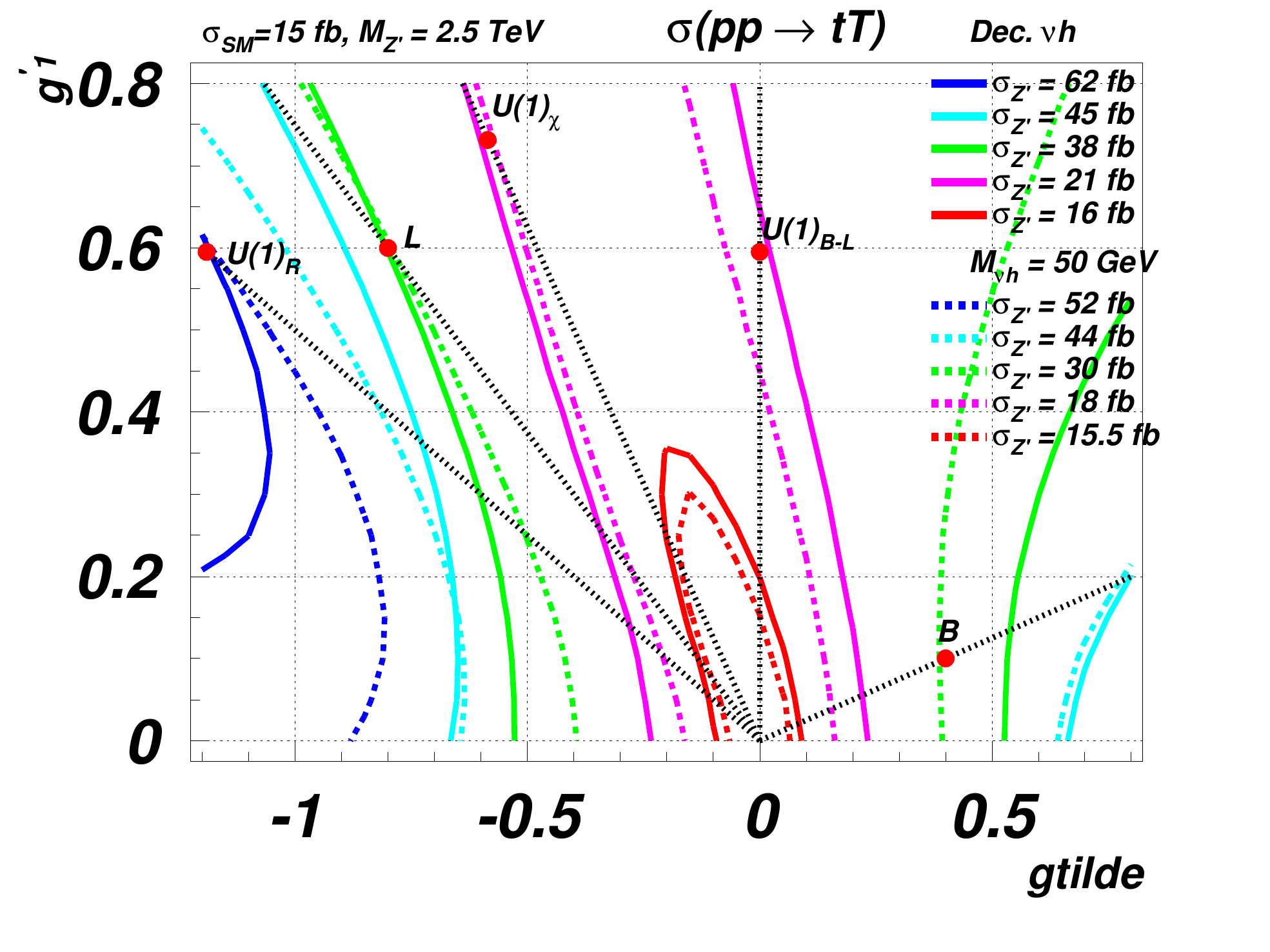}
 \begin{minipage}[b]{0.48\textwidth}
       \hspace{0.05\linewidth}
       \begin{minipage}[b]{0.9\linewidth}      
 \caption{\label{fig:Xs_scan}  Total cross section in fb (signal plus background) on peak $|M_{f\bar{f}}-M_{Z'}|<100$ GeV, for $M_{Z'}=2.5$ TeV for $Z'\to e^+e^-$, $b\bar{b}$ and  $t\bar{t}$, in the ($g'_1$, $\widetilde{g}$) plane, for decoupled neutrino (solid lines) and for $m_{\nu_h}=50$ GeV (dashed lines). The benchmark models are highlighted as (red) dots on the (dotted) scenario lines.}
      \end{minipage}
      \vspace{1cm}
  \end{minipage}
\end{figure}

It is instructive to present at this point the overall event rates in each channel. When including the SM background, the total process reads
\begin{equation}
p,p \to \gamma,\, Z,\, Z' \to f \overline{f}\, .
\end{equation}
To evaluate the total cross sections, the cuts described in eqs.(\ref{cut:peak})--(\ref{cut:bb}) have been applied to enhance the signal. The numerical results are depicted in figure~\ref{fig:Xs_scan}. These are essentially convolutions of the BRs and production cross section detailed in figures~\ref{fig:BRs} and~\ref{fig:total_width} and including small interference effects with the SM background. These figures highlight the different areas of parameter space favoured by each final state and relate directly to the magnitude of the statistical uncertainties in its asymmetries
(to be studied below). 

As previously observed, the total rate when leptons are considered is above the fb level for most of the parameter space. The SM contribution is $0.032$ fb, some orders of magnitude below the signal also when light heavy neutrinos are considered. For the heavy quarks, instead, the SM background is of the same order of the signal, as the former is
largely due to QCD while the latter is (despite being resonant) an EW process. 
We further observe that the total event rate at the LHC for $\sqrt{s}=14$ TeV can be modified when the decay into heavy neutrinos is allowed. In particular, it can diminish by up to $40\%$, $30\%$, and $20\%$ when considering $pp\to\ell^+\ell^-$, $pp\to b\overline{b}$, and $pp\to t\overline{t}$, respectively.

If the presence of light heavy neutrinos lowers the rate for a particular final state, directly increasing its relative error and therefore the error of the asymmetries, we expect this to directly influence the central value of the asymmetries too. Given that the SM background is not altered by the presence of heavy neutrinos, the net effect of a decrease of signal is to increase the relative SM contribution in the samples. Overall, the values of the asymmetries will therefore be more SM-like, i.e., the central values will shift towards the value obtained in the SM. Obviously, this can happen only when the signal-to-background ratio is altered: in the case of leptons, where the SM contribution is negligible, the shift of the central values of the asymmetries is negligible.

Comparing figure~\ref{Zp-excl} to figure~\ref{BRs} and to figure~\ref{fig:Xs_scan}, it is clear that the shape of the current exclusion limits is driven by the BR of the $Z'$ boson into charged leptons: the limits are weaker where the BRs are smaller, i.e., in the region between the $U(1)_\chi$ and the $\notL$ scenarios. It is also clear that searches performed in the top-quark final states do not improve this behaviour, being the BR($Z'\to t\overline{t}$) minimised near  the $U(1)_\chi$ line.  {On the contrary, the $b$-quark final state's BR and cross sections are maximised here, so we checked whether exclusions derived in this final state could improve the overall results. It turns out that the sensitivity in this channel, even for comparable integrated luminosity, i.e., $5$ fb$^{-1}$ as in the analysis of Ref.~\cite{CMS-11-008}, is not comparable to the one exploiting the much cleaner di-lepton final state already discussed, that therefore yields the tighter constraints in the whole parameter space.}


\subsection{Asymmetries}
We have so far described the features of the minimal $Z'$ model in the ($g'_1$, $\widetilde{g}$) plane by analysing standard variables, such as BRs, total width and event rates. For a complete profile of the model, we now move on to study the asymmetries at the $Z'$ peak. Performing a scan over the gauge couplings, the integrated values of asymmetries are presented in the ($g'_{1}$, $\widetilde{g}$) plane for all three final states ($Z^{\prime} \rightarrow \ell^{+}\ell^{-},t\bar{t},b\bar{b}$). The observables were computed using the code described in sect.~\ref{subsec:calc} implementing the cuts in eqs.(\ref{cut:peak})--(\ref{cut:bb}) and folding in the relevant reconstruction efficiencies discussed in sect.~\ref{subsubsec:effs} in order to determine the statistical uncertainties plotted underneath (always for 100 fb$^{-1}$ of integrated luminosity). Finally, we distinguish among light leptons $\ell =e,\mu$, to be used in evaluating $A_{RFB}$, and taus, essential to measure $A_L$ in the leptonic sector.

We present here our results for $A_{RFB}$ and $A_L$ as representatives for charge and spin variables, respectively.
Although we have produced results for it,  we will not present $A_{LL}$, because of its capabilities in the top final state only, while the aim of this paper is to compare predictions in different final states to assess the distinguishability of models.

\subsubsection{Charge asymmetry: $A_{RFB}$}
\begin{figure}[!t]
\centering
\includegraphics[angle=0,width=0.45\textwidth]{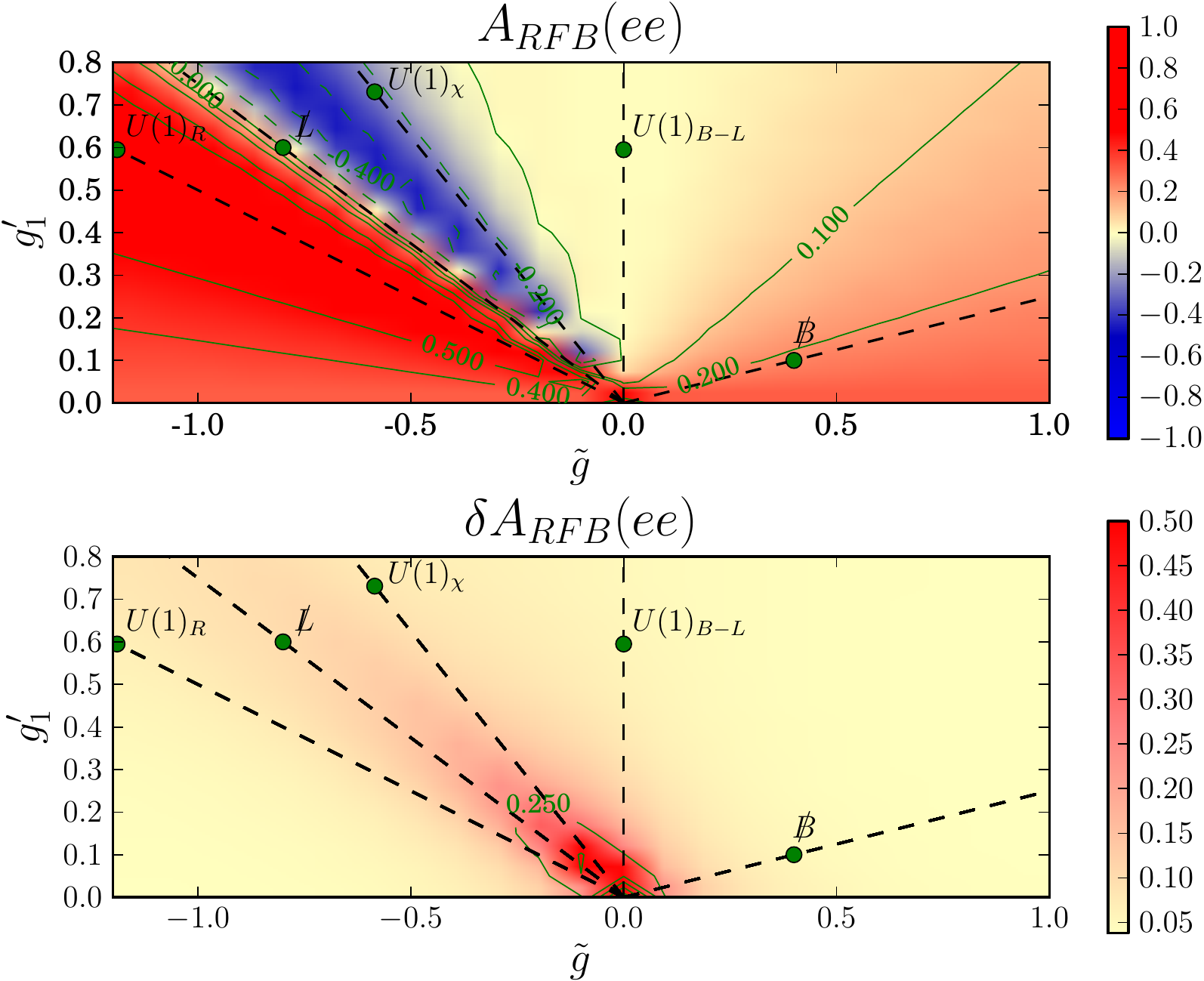}
\includegraphics[angle=0,width=0.45\textwidth]{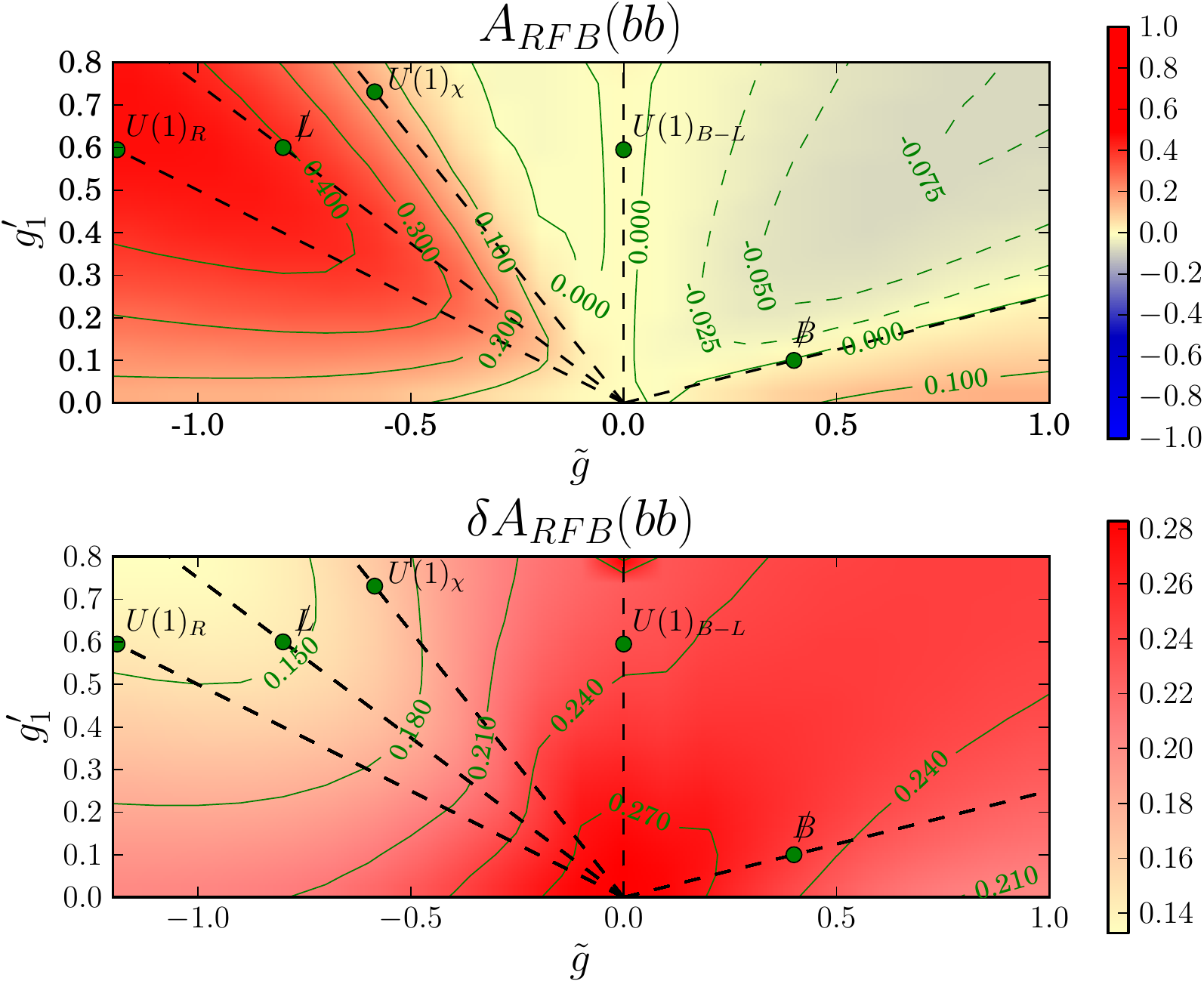}\\
\vspace{1cm}
\includegraphics[angle=0,width=0.45\textwidth]{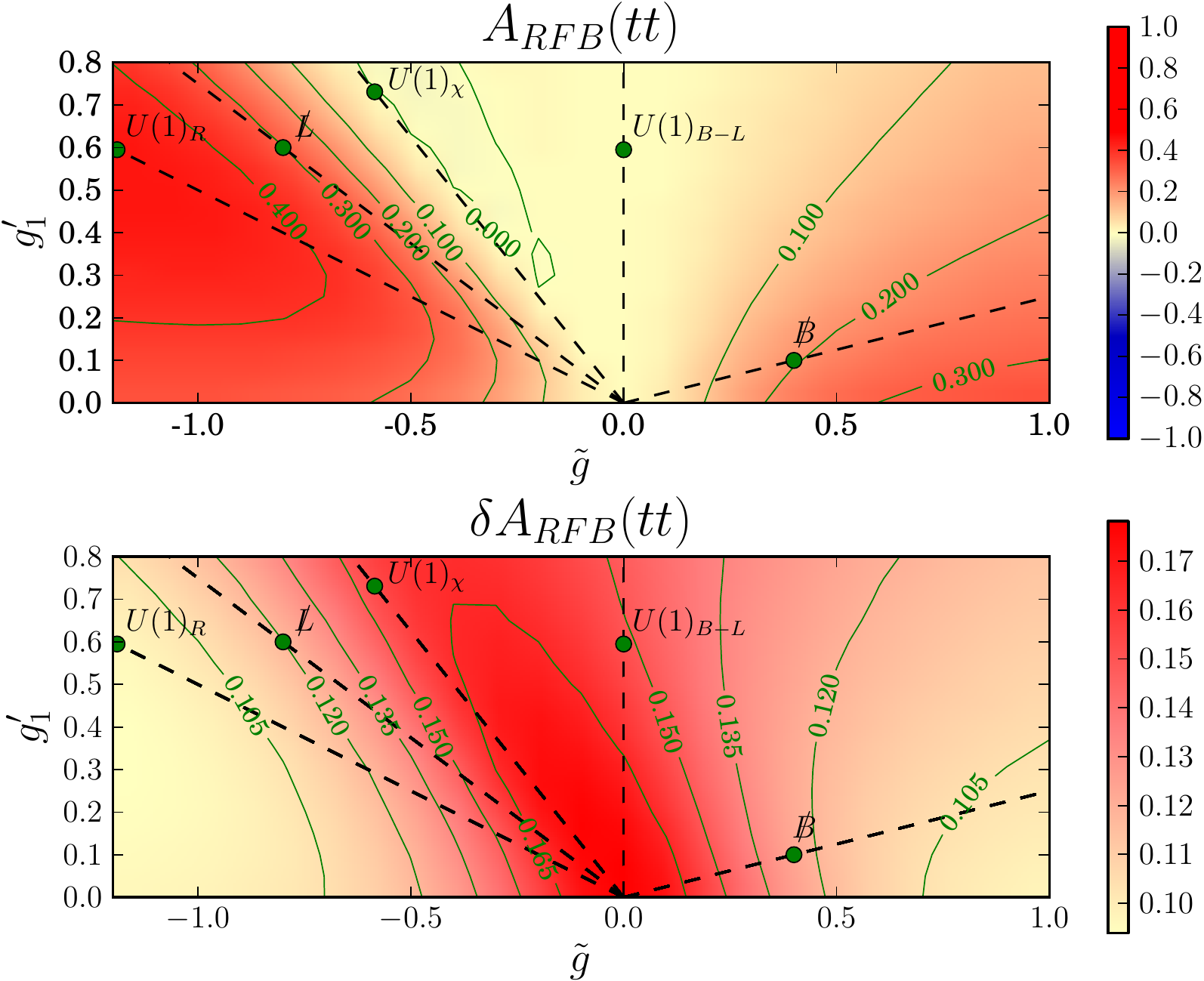} 
\begin{minipage}[b]{0.45\textwidth}
      \hspace{0.05\linewidth}
      \begin{minipage}[b]{0.9\linewidth}     
 \caption{\label{fig:ARFB_scan} $A_{RFB}$ on peak $|M_{f\bar{f}}-M_{Z'}|<100$ GeV, for $M_{Z'}=2.5$ TeV for $Z'\to e^+e^-$, $b\bar{b}$ and  $t\bar{t}$ in the ($g'_1$, $\widetilde{g}$) plane. Statistical uncertainties are shown underneath each plot in the same plane assuming 100 fb$^{-1}$ of integrated luminosity. The benchmark models are highlighted as (green) dots on the (dotted) scenario lines and, where relevant, a contour shows where the asymmetry vanishes.}
     \end{minipage}
 \end{minipage}
\end{figure}

Figure~\ref{fig:ARFB_scan} shows the rapidity dependent forward-backward asymmetry along with its statistical uncertainty in the chosen three final states. In this case, the electron and muon final states can be used and possibly combined to reduce the statistical uncertainty further. One can clearly see the asymmetry vanishing in each final state along the scenario line corresponding to a zero value of one of the chiral couplings (vector or axial), as described in sect.~\ref{sec:benchmarks}.

The magnitude of the statistical uncertainties matches the total cross section plots in figure~\ref{fig:Xs_scan} in accordance with eq.~(\ref{eqn:error}) and, outside of the cases near the trajectories where a particular final state has vanishing asymmetries, the uncertainties can be as low as around 25\% in the case of the top final state. In the leptonic final state, the uncertainties are comparatively smaller, apart form the large spike at the SM point due to the lack of cross section. Finally, the $b\bar{b}$ final state still appears to perform reasonably well with errors of order 30-50\%, considering the extremely low reconstruction efficiency.
The sensitivity to the relative sign of the couplings in the final state is blurred due to the dependence on the product of the initial and final state couplings, as discussed in sect.~\ref{subsubsec:charge}. 

\begin{figure}[!t]
\centering
\includegraphics[angle=0,width=0.45\textwidth]{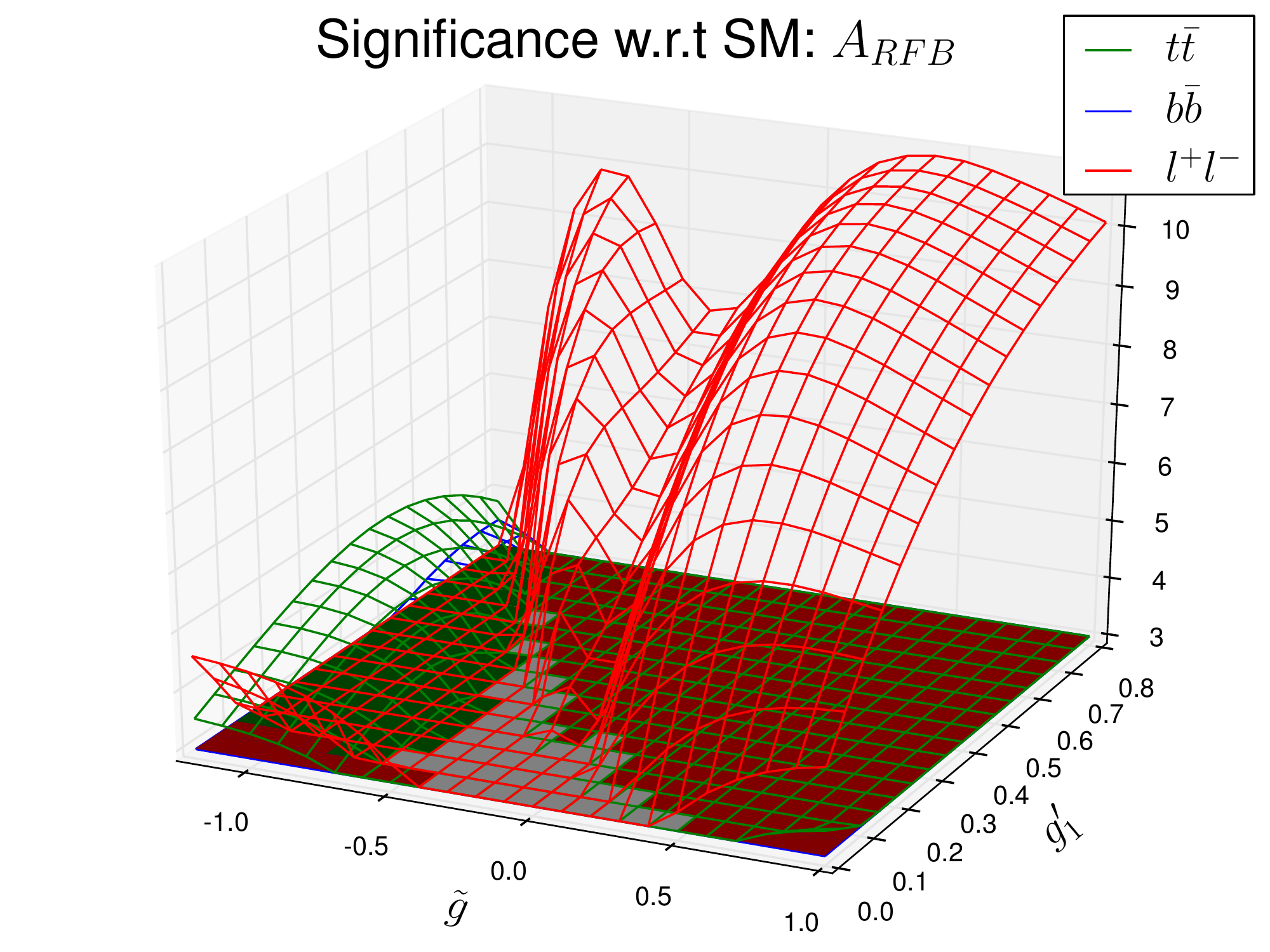}
\includegraphics[angle=0,width=0.45\textwidth]{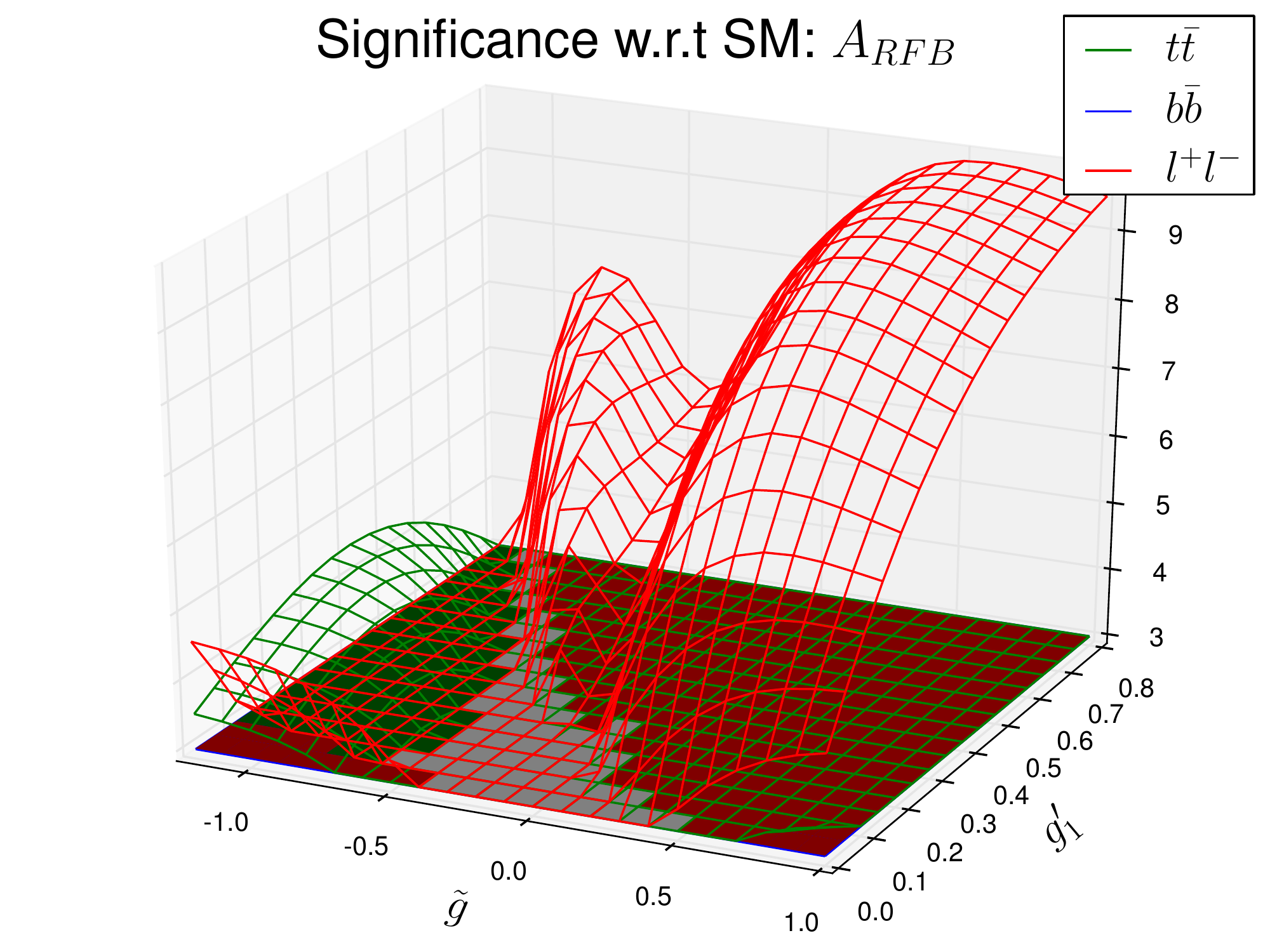} 
 \caption{\label{fig:ARFB_surf} Significance of $A_{RFB}$ with respect to the SM prediction on peak $|M_{f\bar{f}}-M_{Z'}|<100$ GeV, for $M_{Z'}=2.5$ TeV for $Z'\to e^+e^-$ (red), $b\bar{b}$ (blue) and  $t\bar{t}$ (green) in the ($g'_1$, $\widetilde{g}$) plane at the LHC at 14 TeV both for decoupled heavy neutrinos (left) and for heavy neutrinos of 50 GeV (right). The coloured projection denotes the final state which offers the most significance for each pair of coupling values.
We assume here a luminosity of 100 fb$^{-1}$.}
\end{figure}

The overall assessment of the visibility of the $Z^\prime$ boson suffers 
from the large uncertainties (especially if compared to $A_L$, as we will see later).
As shown in figure~\ref{fig:ARFB_surf}, where also the coloured projection labels the final state which offers the highest significance for each pair of coupling values is given, 100 fb$^{-1}$ are enough to gain a sensitivity equal or greater than 3 almost everywhere. The dominance of the leptonic final state is remarkable, this final state being sufficient to cover most of the parameter space. This is due to the large positive value of this asymmetry for the SM, while in our model the leptonic final state yields a value for $A_{RFB}$ much smaller or even slightly negative in most of the parameter space. It is peculiar that, for large negative values of $\widetilde{g}$, where also $A_{RFB}^\ell$ is large, the already small errors are not sufficiently small to allow one to distinguish the $Z'$ boson from the SM. Nonetheless, a good discrimination power is there provided by top quark final states, although with significance always below 5 due to the larger uncertainties.

The effect of introducing the heavy neutrinos is to reduce significances across the board, which results in a slight enlargement of the grey area in which a 3$\sigma$ significance cannot be obtained in any final state. The reason for this is twofold, although driven by a single cause: the reduction of the signal rates in the considered final state. First, as discussed at the end of section~\ref{sec:ev_rates}, the central value of the asymmetry is shifted towards SM values due to a reduced signal-over-SM ratio of events. This does not affect the lepton final state, which significance is reduced due to the second effect: smaller total rates mean larger errors, which naively reduce the significance.

\subsubsection{Polarisation asymmetry: $A_L$}
\begin{figure}[!t]
\centering
\includegraphics[angle=0,width=0.45\textwidth]{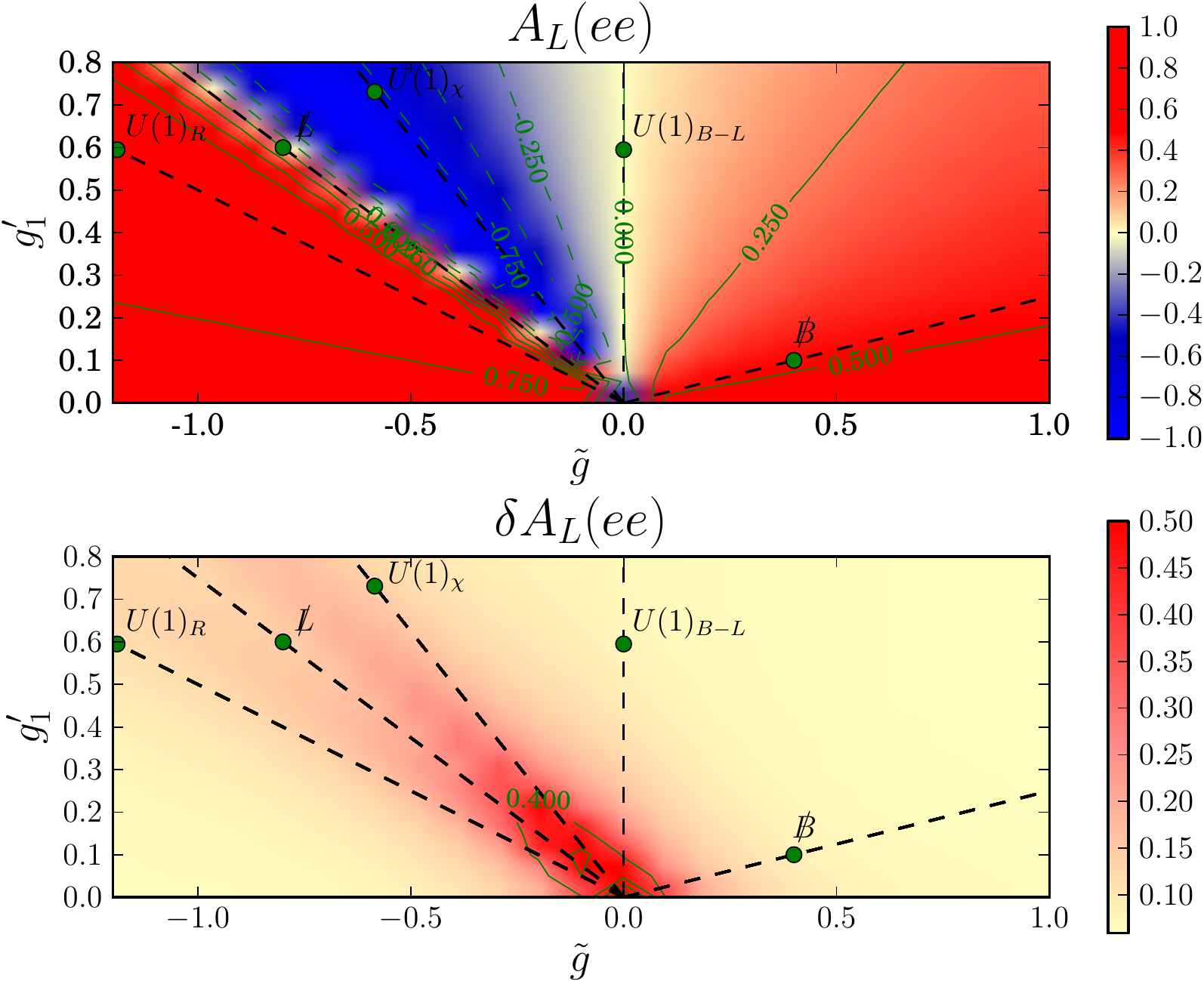}
\includegraphics[angle=0,width=0.45\textwidth]{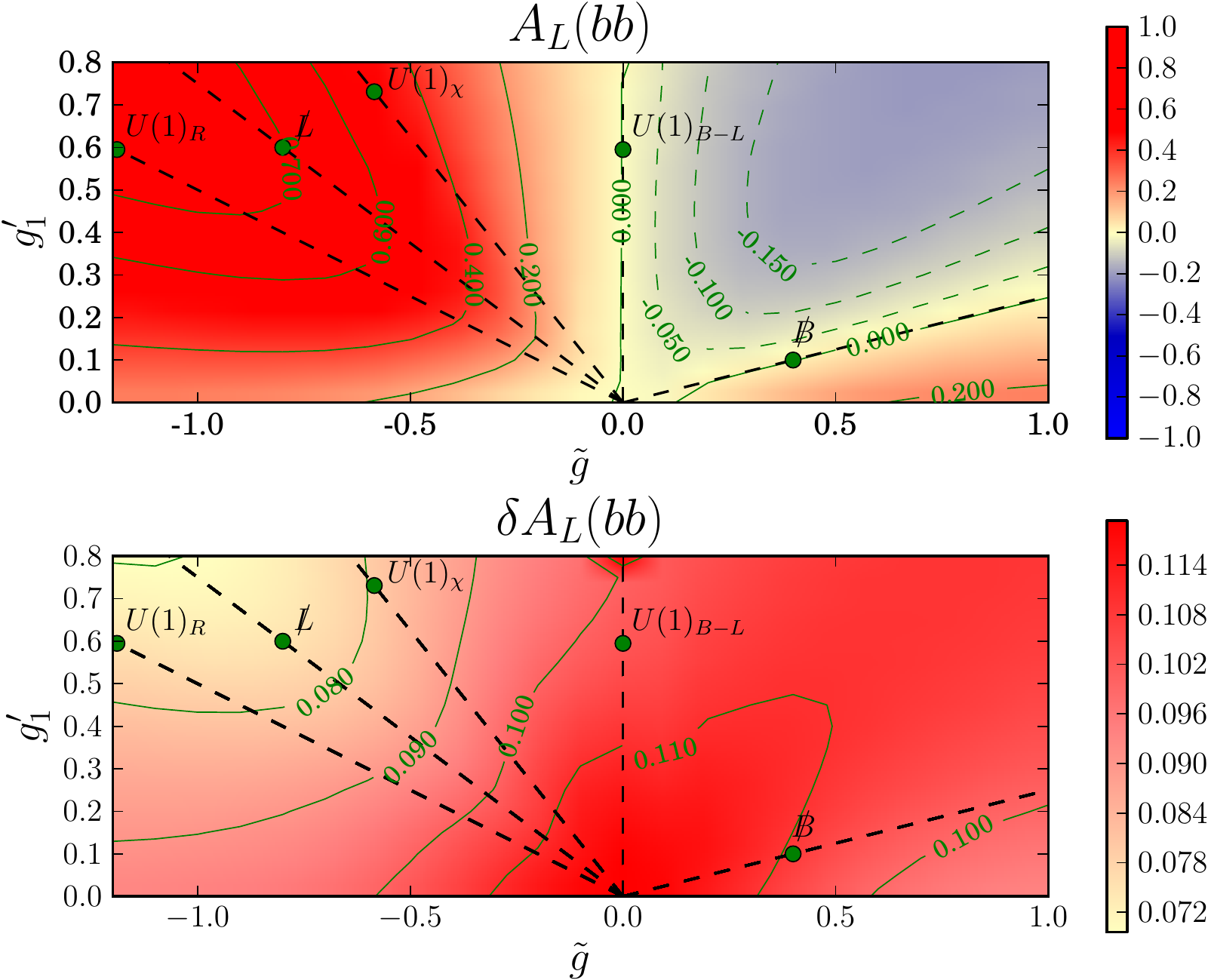}\\
\vspace{1cm}
\includegraphics[angle=0,width=0.45\textwidth]{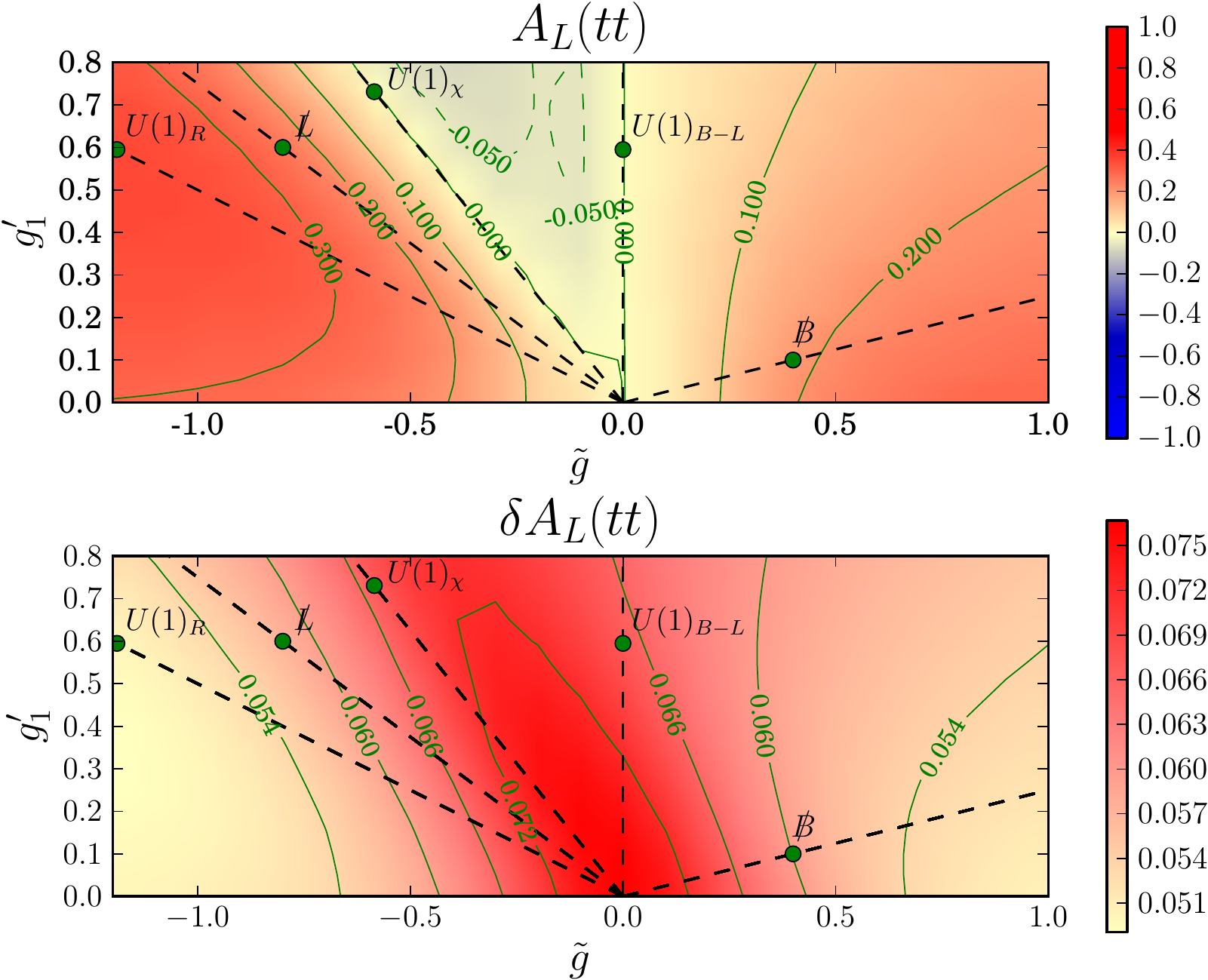}
\begin{minipage}[b]{0.45\textwidth}
      \hspace{0.05\linewidth}
      \begin{minipage}[b]{0.9\linewidth}     
 \caption{\label{fig:AL_scan} $A_{L}$ on peak $|M_{f\bar{f}}-M_{Z'}|<100$ GeV, for $M_{Z'}=2.5$ TeV for $Z'\to e^+e^-$, $b\bar{b}$ and  $t\bar{t}$ in the $g'_1$, $\widetilde{g}$ plane at the LHC at 14 TeV. Statistical uncertainties are shown underneath each plot in the same plane assuming 100 fb$^{-1}$ of integrated luminosity. The benchmark models are highlighted as (green) dots on the (dotted) scenario lines and, where relevant, a contour shows where the asymmetry vanishes.}
     \end{minipage}
 \end{minipage}
\end{figure}

Figure~\ref{fig:AL_scan} shows the single polarisation asymmetry along with its statistical uncertainty in the 
usual three final states. Recall that in the case of the leptonic final state labelled $ee$, the measure would be obtained from the tau final state with the discussed reconstruction efficiency times BR of 10\%.
Once again and even more clearly than for $A_{RFB}$, one can see the asymmetry vanishing in each final state where expected. The change in sign of the asymmetry at different points in parameter space corresponds to a change in relative sign of the vector and axial couplings of the $Z^\prime$ boson to the final state as discussed in sect.~\ref{subsubsec:spin}. This is most pronounced in the di-lepton case because of the comparatively low SM background, which dilutes the asymmetry in the other final states.
The behaviours of the statistical uncertainties do not greatly differ from those in the $A_{RFB}$ plots as they are largely determined by the cross section values, although they are about a factor of two smaller, making this variable the one with the greatest discrimination power. 

\begin{figure}[!t]
\centering
\includegraphics[angle=0,width=0.45\textwidth]{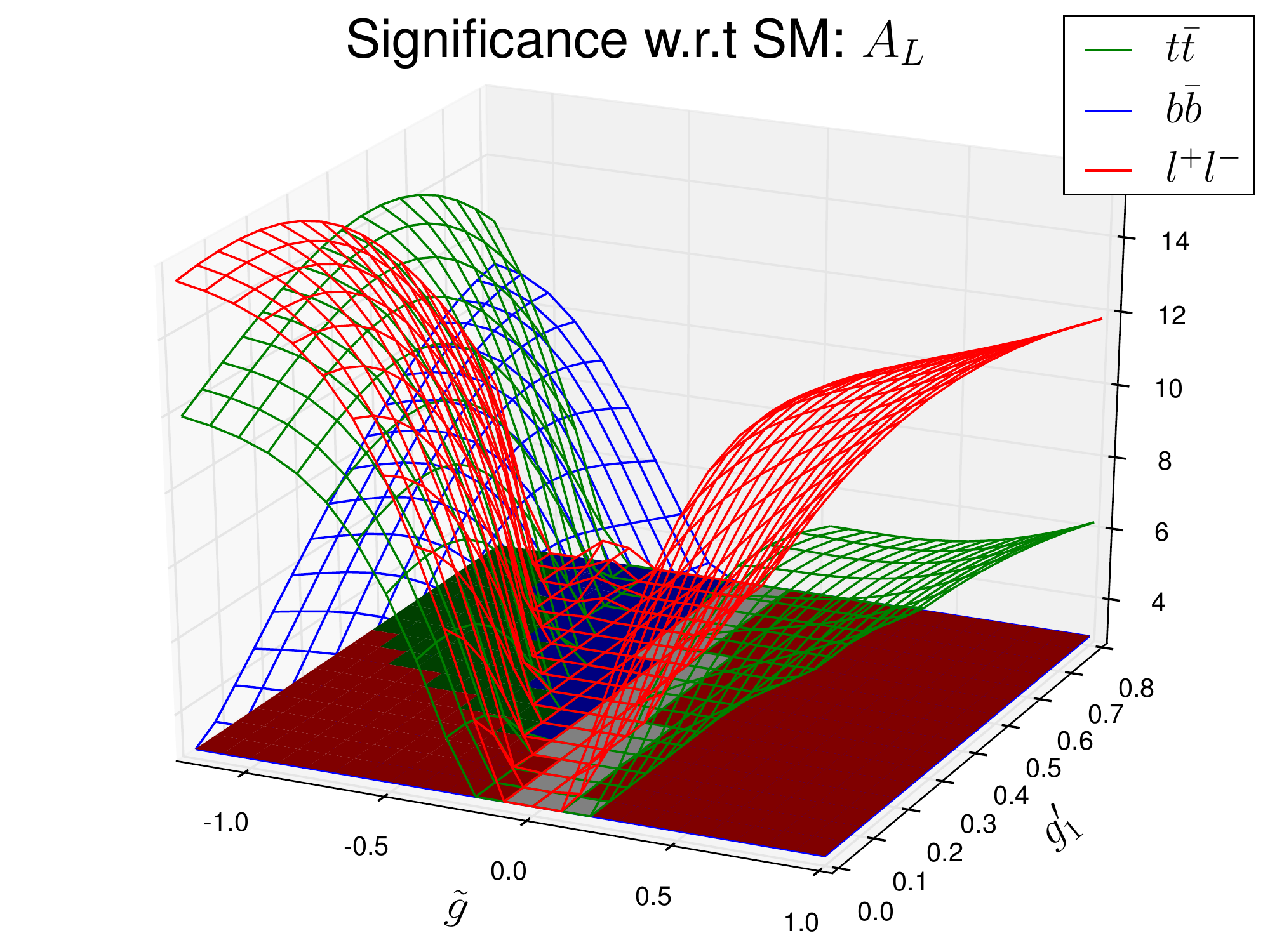}
\includegraphics[angle=0,width=0.45\textwidth]{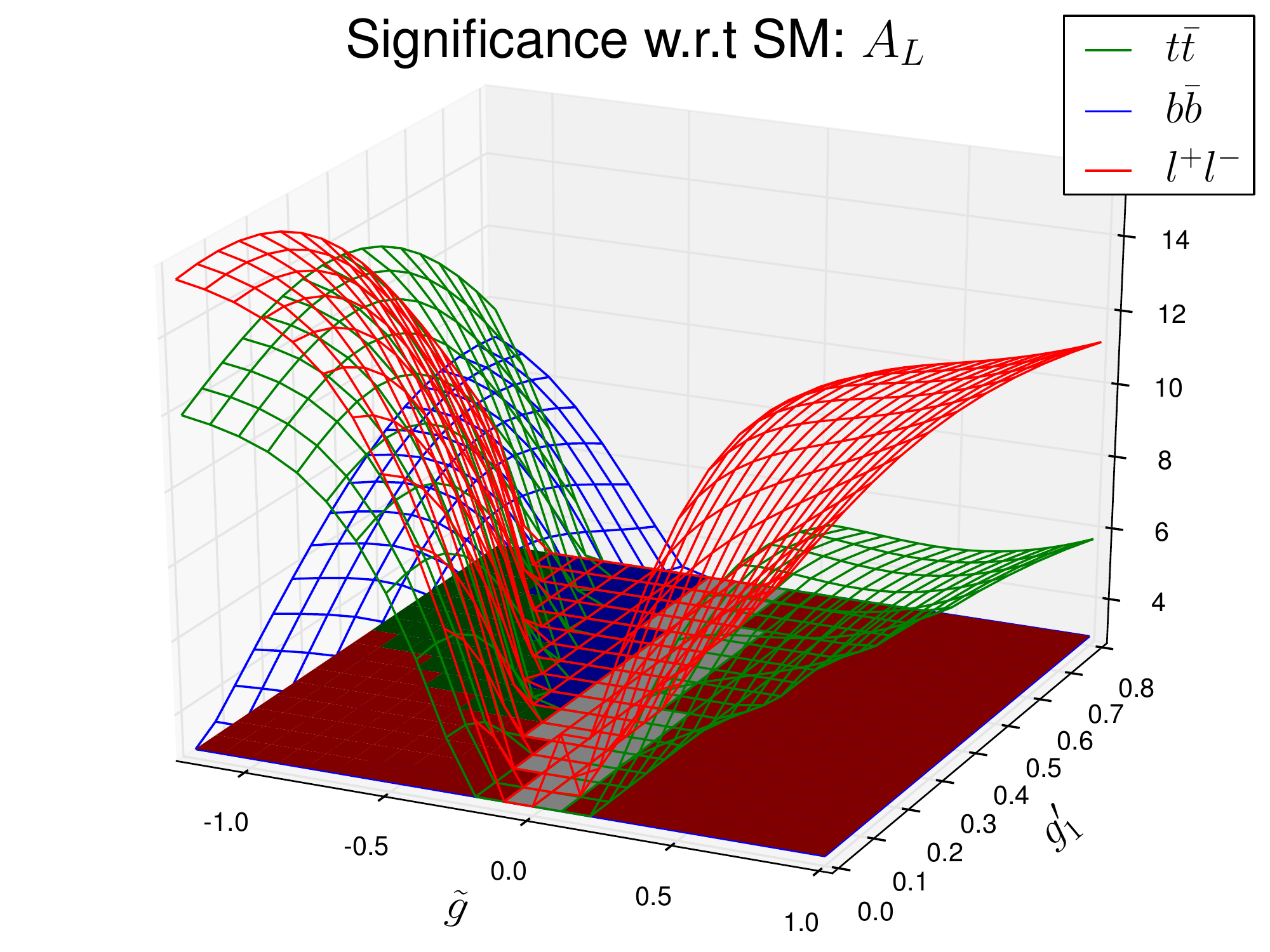}    
 \caption{\label{fig:AL_surf} Significance of $A_{L}$ withe respect to the SM prediction on peak $|M_{f\bar{f}}-M_{Z'}|<100$ GeV, for $M_{Z'}=2.5$ TeV for $Z'\to e^+e^-$ (red), $b\bar{b}$ (blue) and  $t\bar{t}$ (green) in the ($g'_1$, $\widetilde{g}$) plane at the LHC at 14 TeV both for decoupled heavy neutrinos (left) and for heavy neutrinos of 50 GeV (right). The coloured projection denotes the final state which offers the most significance for each pair of coupling values.
We assume here a luminosity of 100 fb$^{-1}$.}
\end{figure}
Overall this gives an indication that all channels could be useful in probing the full parameter space of such a $Z^\prime$ model in complementing one another by having different areas where the sensitivity is the best. This is shown in figure~\ref{fig:AL_surf} where significances in excess of 3$\sigma$ appear over almost all of the  ($g'_1$, $\widetilde{g}$) plane.
Although leptons almost always provide the best significance, due to the large negative SM value as compared to the large positive signal values over the parameter space, in the areas where the two are similar for the lepton final state, $b\bar{b}$ and $t\bar{t}$ come into play and provide adequate `coverage'. 
Again, having coupled heavy neutrinos results in an overall reduction of the signal significance with respect to the SM and in a slight enlargement of the grey area in which 3$\sigma$ significance cannot be obtained in any final state. However, comparing figure~\ref{fig:ARFB_surf} to figure~\ref{fig:AL_surf}, it is clear how these grey areas overlap only in the region where both gauge couplings $g'_1$ and $\widetilde{g}$ are small, showing a great complementarity of observables and final states.

So far we have discussed how well the asymmetry produced by a $Z'$ boson in our model can be distinguished from the SM background. One would also like to  disentangle different combinations of gauge couplings, in turn leading to their absolute measurement. Although the full analysis and especially this last part are beyond the scope of this paper, we would like to briefly discuss the first point.

\begin{figure}[!t]
\centering
\includegraphics[angle=0,width=0.45\textwidth]{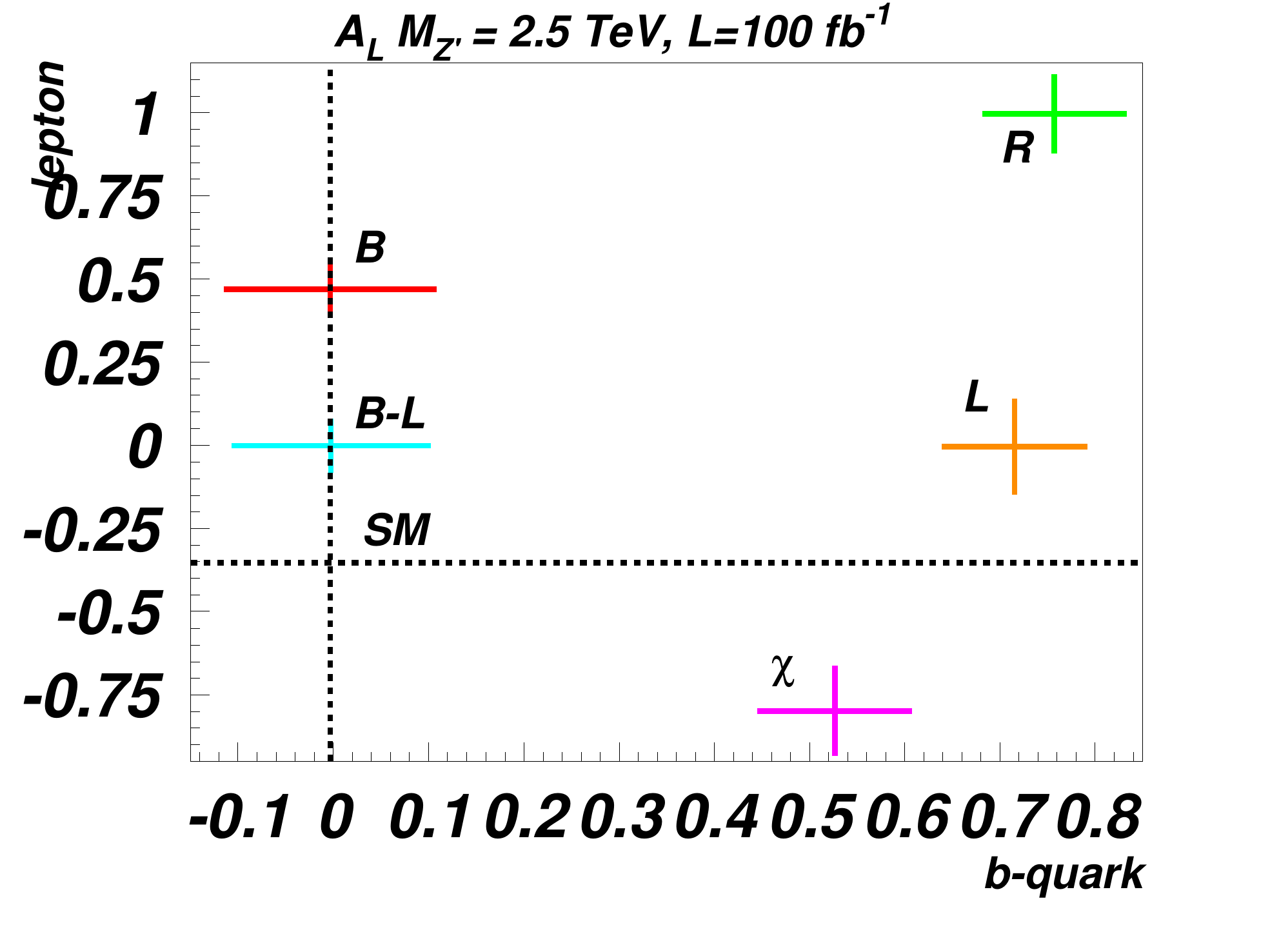}
\includegraphics[angle=0,width=0.45\textwidth]{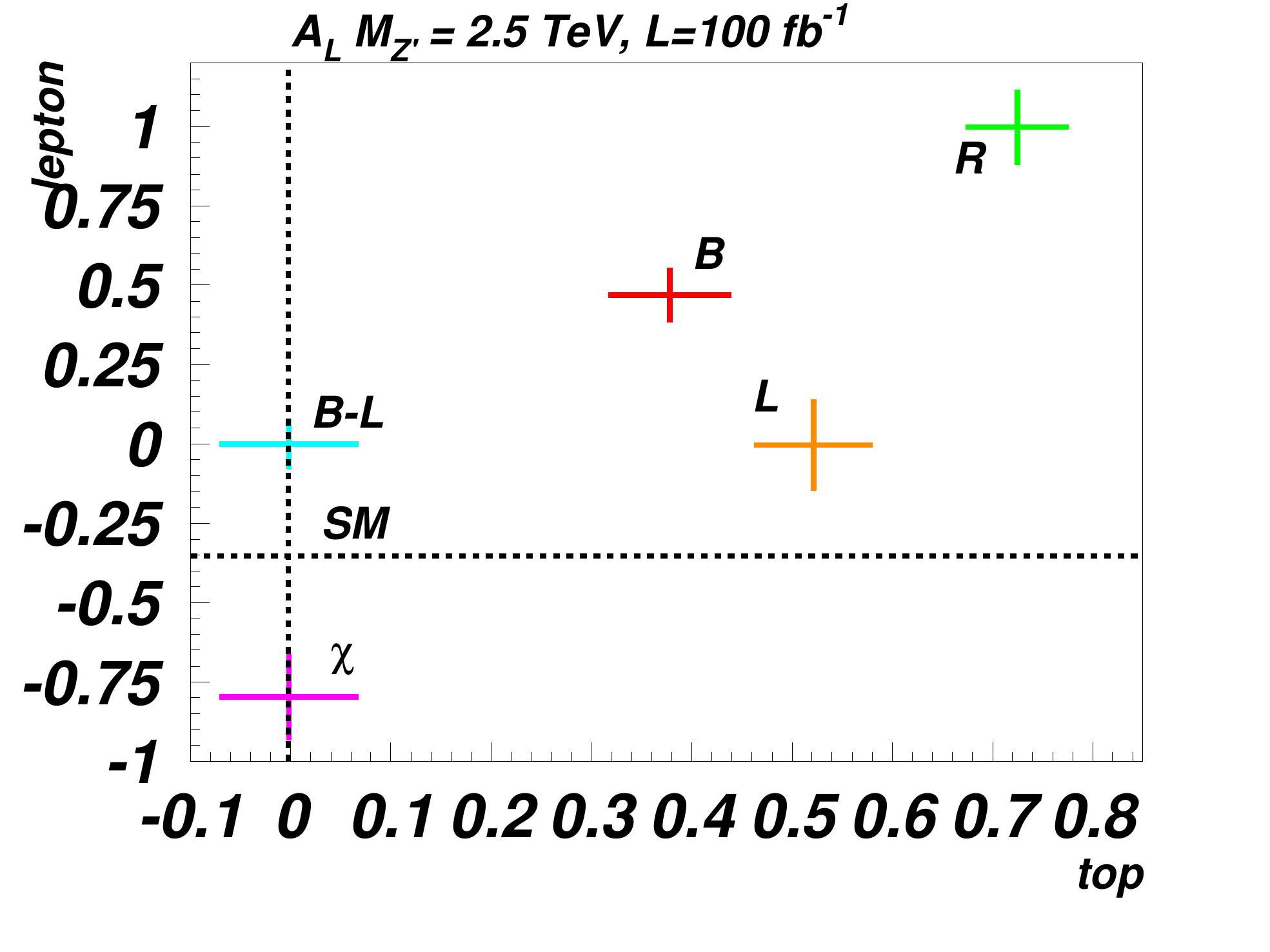}\\
\vspace{1cm}
\includegraphics[angle=0,width=0.45\textwidth]{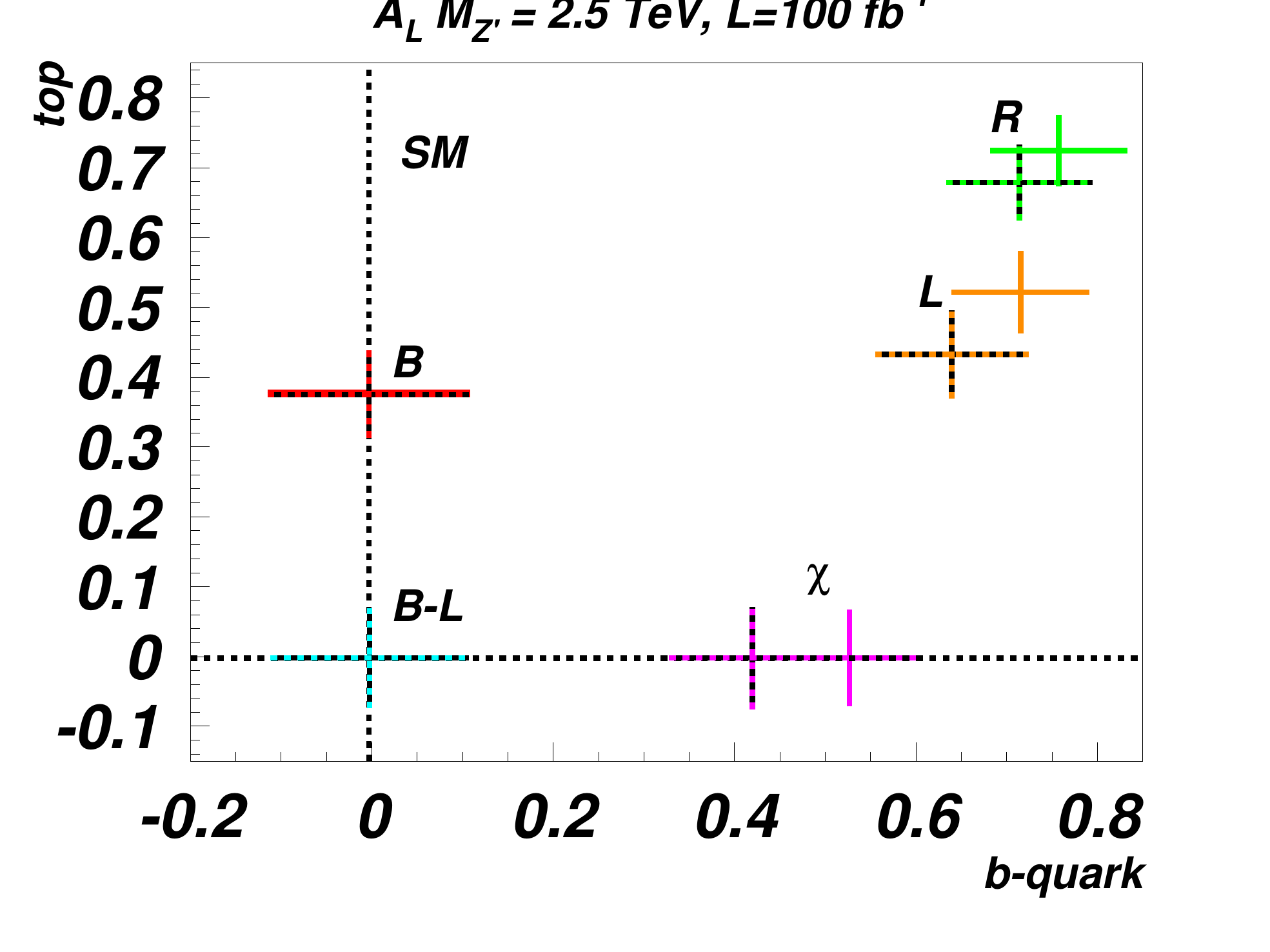} 
\begin{minipage}[b]{0.45\textwidth}
      \hspace{0.05\linewidth}
      \begin{minipage}[b]{0.9\linewidth}     
 \caption{\label{fig:AL_comp} On-peak $A_{L}$ values with relative errors for $M_{Z'}=2.5$ TeV in different combinations of final states, for the selected benchmark models in sect.~\ref{sec:benchmarks}. Solid crosses are for decoupled neutrinos, while dashed crosses are for $m_{\nu_h}=50$ GeV. Leptons do not show any appreciable difference, hence they are not displayed.}
     \end{minipage}
 \end{minipage}
\end{figure}

Figure~\ref{fig:AL_comp} shows that all our benchmark models can well be discriminated with 100 fb$^{-1}$ of data. Especially, degenerate models in one final state (such as $B-L$ and $\chi$ in the case of top quarks) are well separated either using leptons or, eventually, $b$-quarks. The last frame, displaying the $t$- versus $b$-quark case, clearly shows the impact of including heavy neutrinos: represented by dashed crosses, the asymmetries with light $\nu_h$ are closer to the SM values and have larger errors.

\begin{table}[h!]
	\centering
\scalebox{0.9}{
	\begin{tabular}{|c|ccc|ccc|ccc|ccc|ccc|ccc|}\hline
	$A_L$      & \multicolumn{3}{c|}{$\notB$} &\multicolumn{3}{c|}{$R$}&\multicolumn{3}{c|}{$\notL$}&\multicolumn{3}{c|}{$B-L$}&\multicolumn{3}{c|}{$\chi$}\\ 
50 $\backslash$ dec&$\tau$ & $t$ &$b$ &$\tau$ & $t$ &$b$ &$\tau$ & $t$ &$b$ &$\tau$ & $t$ &$b$ &$\tau$ & $t$ &$b$ \\ \hline
 $\notB$ & \multicolumn{3}{c|}{--}      & 3.6&4.3&5.6 & 2.8&1.7&5.3 & 4.0&4.1&0 & 7.9&4.1&3.2  \\
 $R$     & 3.6&3.7&5.2 & \multicolumn{3}{c|}{--} & 5.4&2.6&0.4 & 7.0&8.4&5.9 & 10.0&8.4&1.6\\
 $\notL$ & 2.8&0.7&4.6 & 5.4&2.9&0.6 & \multicolumn{3}{c|}{--} & 0&5.7&5.5 & 4.0&5.7&1.3\\
 $B-L$   & 4.0&4.0&0 & 7.0&7.6&5.3& 0&4.5&4.6 & \multicolumn{3}{c|}{--} & 5.0&0&3.3\\
 $\chi$  & 7.9&3.9&2.9 & 10.0&7.5&2.4 & 4.0&4.5&1.8 & 5.0&0&3.0 & \multicolumn{3}{c|}{--} \\
	\hline
	\end{tabular}}
\caption{Significance for $A_L$ for the LHC at 14 TeV for $100$ fb$^{-1}$ and $M_{Z'}=2.5$ TeV for the common benchmark points in the $\tau,t,b$ final states. Upper triangle for decoupled heavy neutrinos and lower triangle for $m_{\nu_h}$=50 GeV.}\label{tab:signif_LHC14_AL}
\end{table}

{
In table~\ref{tab:signif_LHC14_AL} we collect the significance of each benchmark point with respect to any other when $A_L$ is measured in each of the final states, would the measure in $b$-quark final state be available, both for decoupled neutrinos (upper triangle) and for $m_{\nu_h}=50$ GeV (lower triangle). We first notice that most of the models are quite distinguished from each other just by looking at taus, that generally speaking delivers higher discrimination power due to the smaller errors. The only exception is the  separation between $\notL$ and $B-L$, both with vanishing asymmetries in the lepton final state. Here, the supplementary information from the top final state already proves to be sufficient to discriminate among all the models. If the inclusion of tops already helps to disentangle the models, it is clear that were it available, the measure in $b$-quarks would be of great help to fully distinguish them: the discrimination power is always above the $4\sigma$ level in at least one final state, the $b$-quarks being especially relevant to discriminate between $\notL$ and $\notB$.

Regarding the inclusion of heavy neutrinos, their impact is that of reducing the absolute value of the asymmetry and to increase its statistical error.
Taus are not affected by their presence and therefore show the same significance. As well, the absolute values of the asymmetries for $B-L$ and $\notB$ are not affected by heavy neutrinos, given that the former has vanishing asymmetries everywhere and that the latter has negligible BRs into them. Therefore, their disentanglement from any other model can only get worse, due to the reduced central values of the latter. The same is true also for $\chi$ when tops are considered, given that its asymmetries in this final state are vanishing. When instead models with finite values of the polarisation asymmetry in a specific final state are compared (i.e., $R$ and $\notL$ using tops and $R$, $\notL$ and $\chi$ using $b$'s), the discrimination power therein gets enhanced when considering coupled heavy neutrinos. This seems counterintuitive, but it can be understood looking at figure~\ref{fig:BRs} in terms of BRs. In fact,  the central values change differently and the various models considered here are better separated, despite the slightly larger statistical errors.
}

\begin{figure}[!ht]
\centering
 \includegraphics[angle=0,width=0.49\textwidth ]{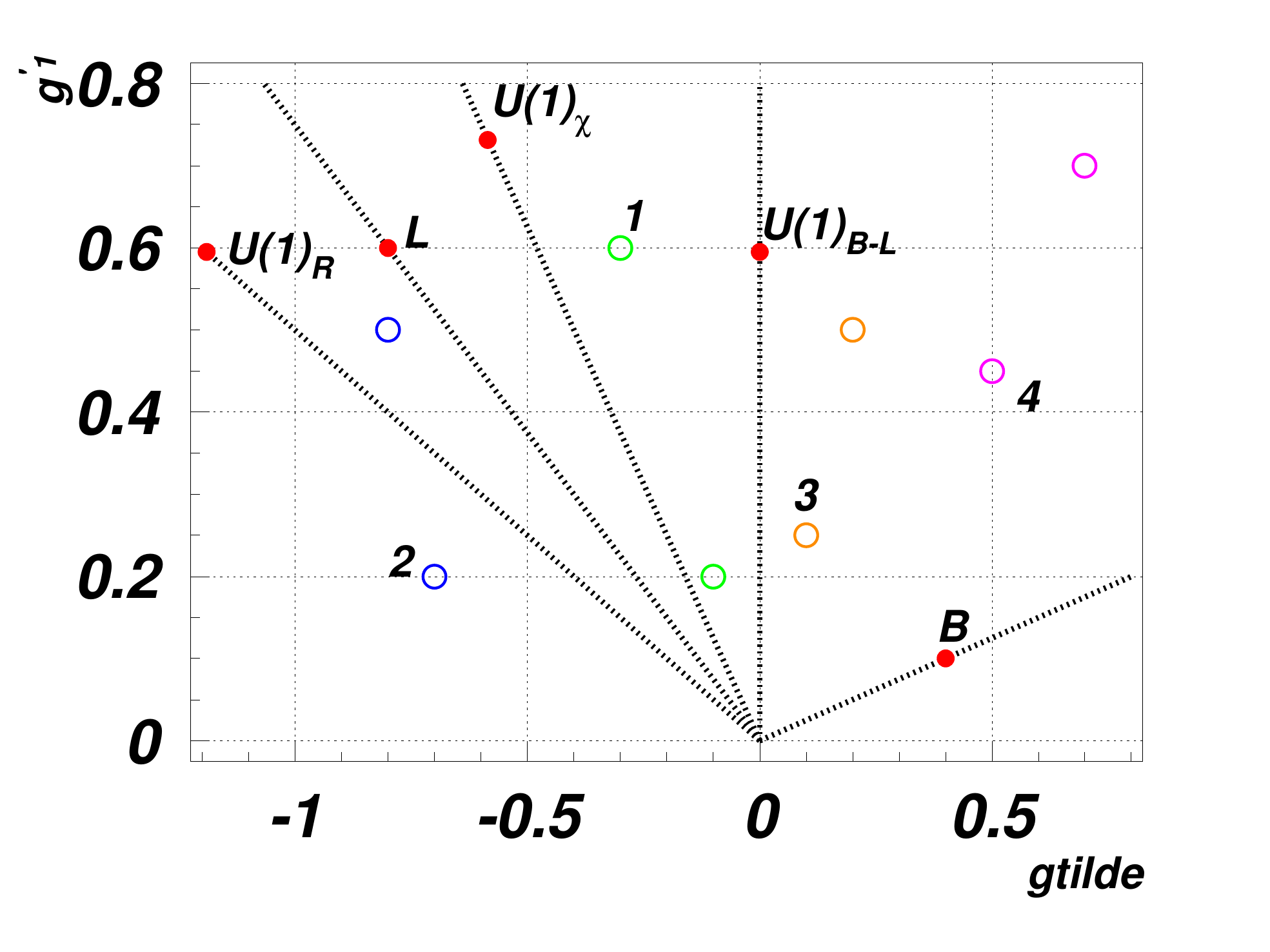}
 \includegraphics[angle=0,width=0.49\textwidth ]{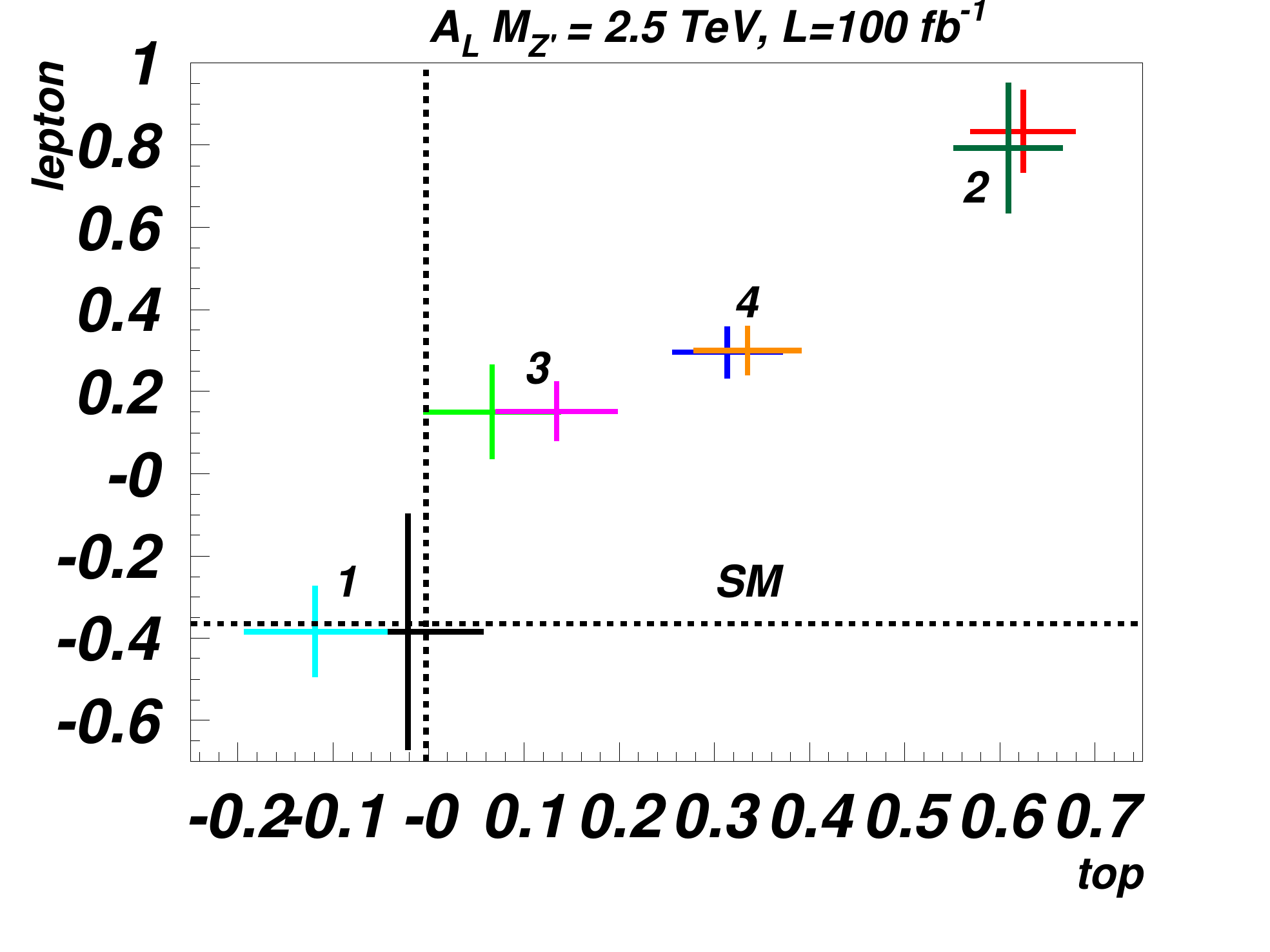}\\
 \includegraphics[angle=0,width=0.49\textwidth ]{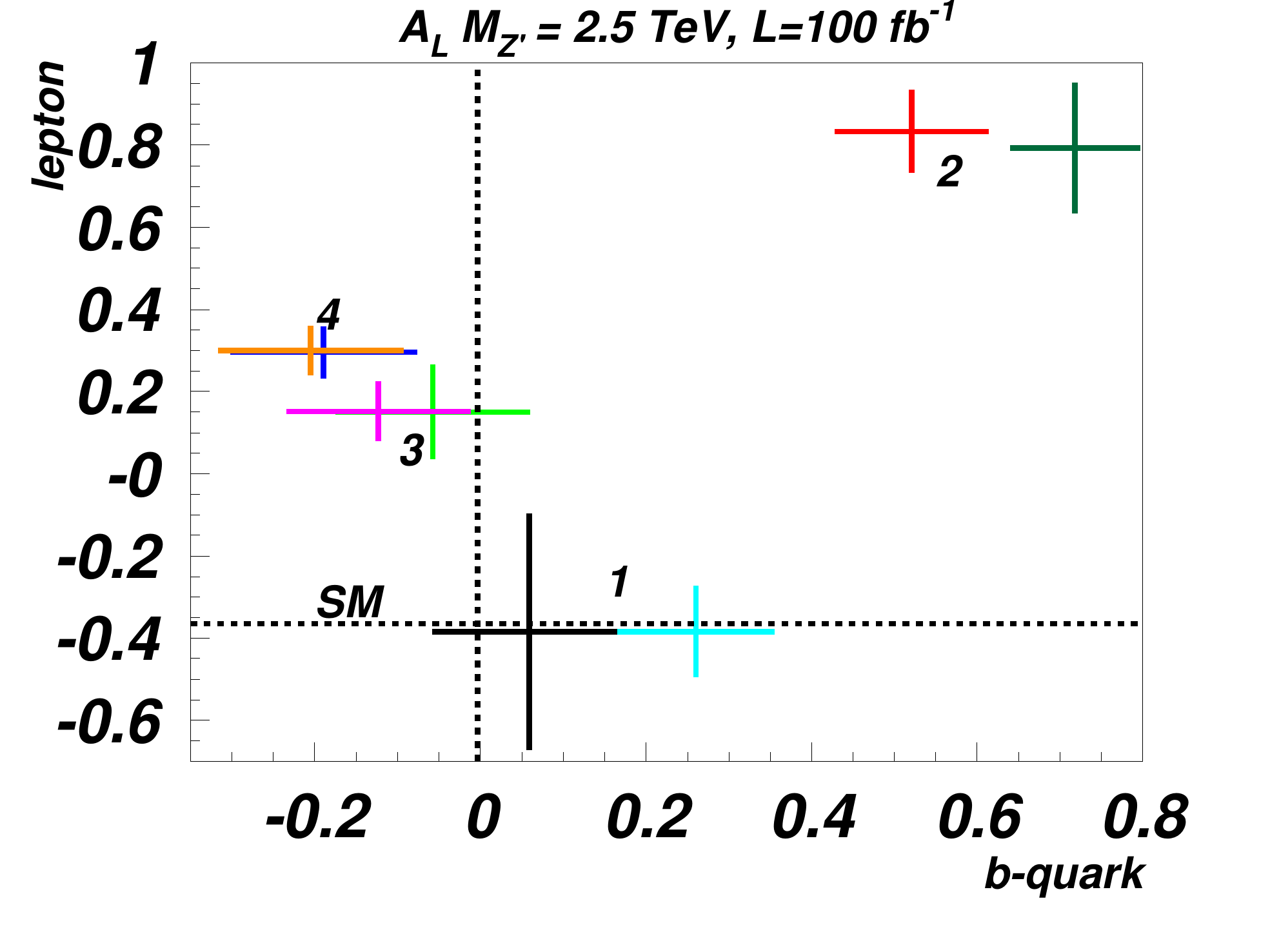}
 \includegraphics[angle=0,width=0.49\textwidth ]{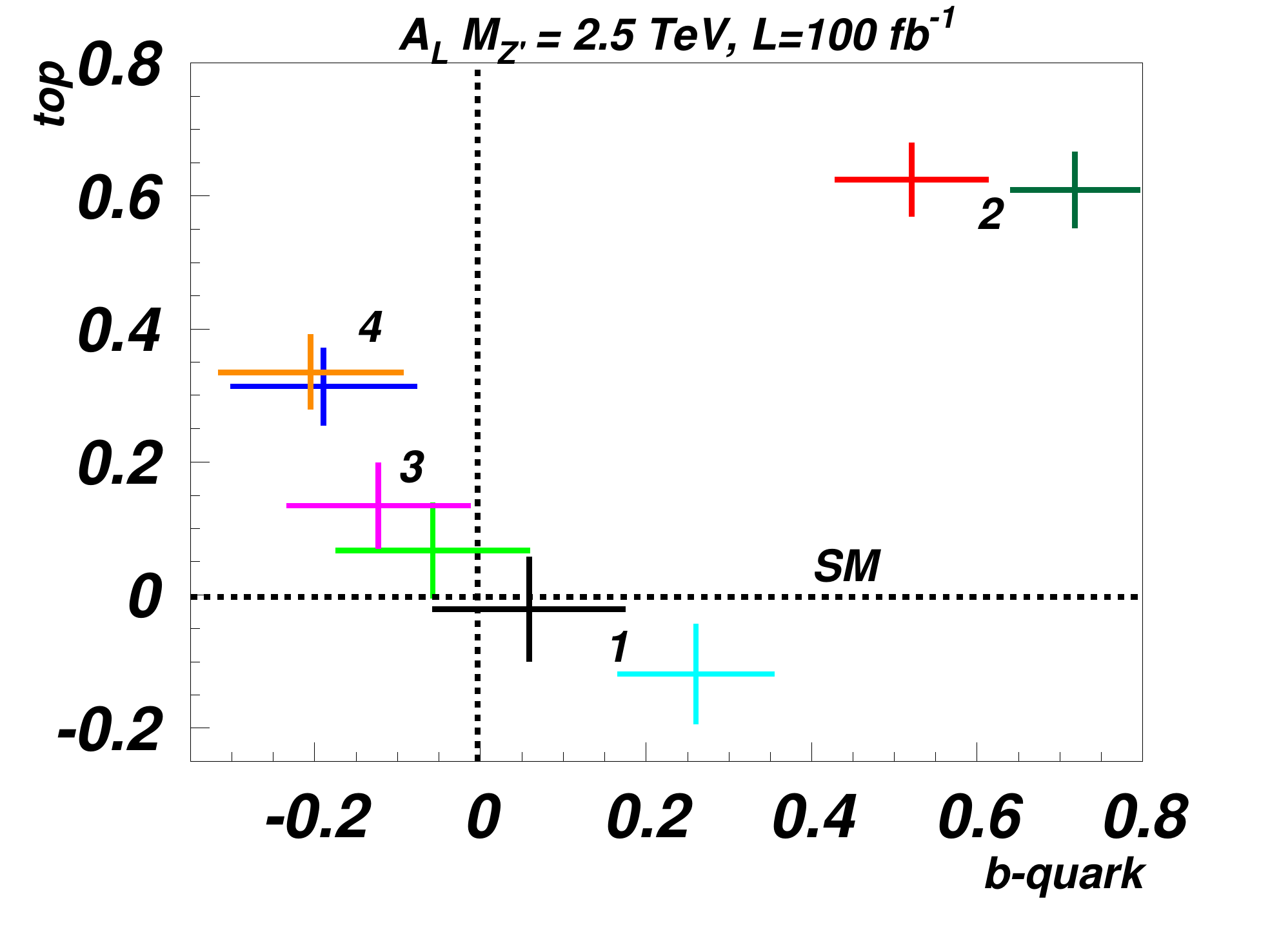}\\
 \caption{\label{fig:AL_fstdeg} (Top-left) combinations of gauge couplings with some degenerate asymmetries. (Other frames) On-peak $A_L$ values with relative errors for combinations of final states for $M_{Z'}=2.5$ TeV and decoupled heavy neutrinos.}
\end{figure}

As a last example, we want to make the case of whether all possible gauge coupling combinations could be resolved in our approach.
{It is not hard to see in figure~\ref{fig:AL_surf} that there exist combinations of couplings that yield similar (within errors) asymmetries in top and tau final states. Possibly, the study of differential and angular distributions of charged leptons, as in Ref.~\cite{Petriello:2008zr}, can fully resolve the whole parameter space. However, here we would like to make a different point: if the measure of the (spin) asymmetries in $b$-quark final states becomes available, it would be of great interest.}
To this purpose, we select four pairs of gauge couplings, shown in figure~\ref{fig:AL_fstdeg}(top-left) that have degenerate asymmetries both for leptons and for tops (see fig.~\ref{fig:AL_fstdeg}(top-right)). The pairs of points labelled ``1'' and ``2'' are examples of combinations that could be discriminated when the asymmetry in $b$-quarks is also considered. However, this is still not possible over the whole parameter space, as clear from the points labelled ``3'' and ``4'', fully degenerate, so in any case one would require to integrate such information with a study of standard observables such as (differential) cross sections, lineshape fits and BRs, reinforcing the importance of making use of all accessible information. We remark that these last statements are valid at 100 fb$^{-1}$ and for the efficiency factors employed. Higher amounts of data, and better efficiencies, will certainly improve the discrimination power, {as well as new generations of micro-vertex detectors could make the $b$-quark final state competitive with the $\tau$-lepton and $t$-quark ones.}

%% file: sect4.tex
In summary, we have proven the feasibility of profiling a $Z'$ boson, possibly discovered at the 14 TeV LHC,
 thanks to the exploitation of the decays of the new gauge boson into $\tau^+\tau^-$, $t\bar t$ {and, possibly,} $b\bar b$ 
states, with respect to the time-honoured studies of DY channels only (i.e., decays into $e^+e^-$ and
$\mu^+\mu^-$ final states). In fact, the former signatures afford one with the possibility of not only defining standard charge asymmetries,
the only ones accessible in the latter, but also spin ones. In essence, such heavy fermions (i.e., $\tau$, $b$ and $t$ states) can transmit 
their spin properties to their decay products rather efficiently
(even though just in some specific cases),
thereby making it possible to {define} single- and/or double-spin asymmetries that 
are peculiarly dependent on the quantum numbers of the new gauge boson. 
{Their measure is certainly feasible for tau and top-quark final states, while, at this moment in time, the $b$-quark final state may prove difficult, if not possible at all, to analyse due to experimental limitations in reconstruction, as well as to the highly boosted kinematics.}
Further, based on a dedicated parton level simulation, yet including 
some selection criteria and realistic detector efficiencies, we have argued that {in our approach} neither DY nor any of the above new channels can be used 
alone to fully {probe all the parameter space}. 
{Rather, one way to render this possible is to combine two or more of these channels, including also spin observables,}
as the aforementioned final states show different sensitivity to the observables studied. {Nonetheless, due to the continuous nature of the parameter space, degenerate regions still exist.} We finally made also the point on how the presence of light extra states (the heavy neutrinos) can alter the observables under study, and that such impact could be resolved.
Such conclusions can be extended to a large variety
of the most popular $Z'$ models studied nowadays, all compliant with current experimental limits, and in presence of SM irreducible background 
effects. 

{Ultimately, the goal of profiling an observed resonance would be to measure all the parameters of its fermionic interactions. In the minimal case, as discussed in section 2, there are $5$ independent couplings. Following Ref.~\cite{Petriello:2008zr}, at least one more linearly independent observable is required than those available in the light lepton sector, where some degeneracy still ocours. Considering multiple final states and their spin asymmetries could provide the necessary measurements to extract the fermionic couplings in this minimal case, that become essential when testing the univerality of couplings or when considering models with more general coupling structures. Finally, our continuous scanning of the parameter space highlighted regions where leptonic final states are not the most sensitive ones, hence where alternative final states would prove to be more effective.
}